\documentstyle[12pt]{article}

\def\Z {{\bf Z}}
\def\R {{\bf R}}
\def\C {{\bf C}}

\newcommand{\bra}[1]{\left\langle\, #1\,\right|}
\newcommand{\ket}[1]{\left|\, #1\,\right\rangle}
\newcommand{\bracket}[2]{\left\langle\, #1\,|\,#2\,\right\rangle}
\newcommand{\VEV}[1]{\left\langle #1\right\rangle}
\newcommand{\wt}{\widetilde}
\newcommand{\wh}{\widehat}
\newcommand{\ol}{\overline}
\newcommand{\ra}{\rightarrow}

\newcommand{\nn}{\nonumber}

\newcommand{\cA}{{\cal A}}

\newcommand{\cD}{{\cal D}}
\newcommand{\cH}{{\cal H}}
\newcommand{\cF}{{\cal F}}
\newcommand{\cO}{{\cal O}}

\newcommand{\bX}{{\bf X}}
\newcommand{\bP}{{\bf P}}
\newcommand{\bx}{{\bf x}}

\newcommand{\bz}{{\bf z}}
\newcommand{\bp}{{\bf p}}
\newcommand{\bG}{{\bf\Gamma}}
\newcommand{\bM}{{\bf M}}
\newcommand{\Dslash}{D\!\!\!\!\slash\,}

\newcommand{\del}{\partial}

\newcommand{\half}{\frac{1}{2}}

\newcommand{\AD}[1]{$\ol{\mbox{D~\,}}\!\!\!$#1}

\newcommand{\beq}{\begin{eqnarray}}
\newcommand{\eeq}{\end{eqnarray}}
\newcommand{\beqa}{\begin{eqnarray}}
\newcommand{\eeqa}{\end{eqnarray}}
\newcommand{\CR}{\nonumber \\}

\def\tr{\mathop{\rm tr}\nolimits}

\def\Tr{\mathop{\rm Tr}\nolimits}
\def\TrP{\mathop{\rm Tr\,P}\nolimits}
\def\trP{\mathop{\rm tr\,P}\nolimits}
\def\Str{\mathop{\rm Str}\nolimits}
\def\TrwhP{\mathop{\rm Tr\,\wh P}\nolimits}
\def\trwhP{\mathop{\rm tr\,\wh P}\nolimits}

\def\mod{\mathop{\rm mod}\nolimits}

\makeatletter
\@addtoreset{equation}{section}
\makeatother

\def\mat#1{\matt[#1]}
\def\matt[#1,#2,#3,#4]{\left(%
\begin{array}{cc} #1 & #2 \\ #3 & #4 \end{array} \right)}

\def\v2#1{\vv2[#1]}
\def\vv2[#1,#2]{\left(%
{#1 \atop #2}\right)}

\def\drawbox#1#2{\hrule height#2pt
        \hbox{\vrule width#2pt height#1pt \kern#1pt
              \vrule width#2pt}
              \hrule height#2pt}

\def\Fund#1#2{\vcenter{\vbox{\drawbox{#1}{#2}}}}
\def\Asymm#1#2{\vcenter{\vbox{\drawbox{#1}{#2}
              \kern-#2pt       
              \drawbox{#1}{#2}}}}

\def\fnd{\Fund{5.5}{0.4}}
\def\asym{\Asymm{5.5}{0.4}}
\def\sym{\fnd\kern-0.4pt\fnd}

\begin{document}

\begin{titlepage}
\vspace*{-2.5cm}
\null
\begin{flushright}
hep-th/0212188 \\
YITP-02-62\\
ITFA-2002-39\\
December, 2002
\end{flushright}
\vspace{0.5cm}
\begin{center}
{\large \bf
Exact Description of D-branes via Tachyon Condensation
\par}
\lineskip .75em
\vskip1.5cm
\normalsize

{\large
 Tsuguhiko Asakawa\footnote{
E-mail:\ \ {\tt asakawa@yukawa.kyoto-u.ac.jp} },
 Shigeki Sugimoto\footnote{
E-mail:\ \ {\tt sugimoto@nbi.dk}}  and
Seiji Terashima\footnote{
E-mail:\ \ {\tt sterashi@science.uva.nl}} }
\vskip 2.5em
{
${}^1$
\it
Yukawa Institute for Theoretical Physics,
Kyoto University,\\
Kyoto 606-8502, Japan
}
\vskip 1em
{
${}^2$
 \it
The Niels Bohr Institute,\\
Blegdamsvej 17, DK-2100 Copenhagen \O, Denmark \\}
\vskip 1em
{
${}^3$
 \it  Institute for Theoretical Physics,
University of Amsterdam \\
Valckenierstraat 65,
1018 XE, Amsterdam, The Netherlands}
\vskip 2em

{\bf Abstract}

\end{center}

We examine the fluctuations
around a D$p$-brane solution
in an unstable D-brane system using boundary states
and also boundary string field theory.
We show that the fluctuations
correctly reproduce the fields
on the D$p$-brane. Plugging these into the action
of the unstable D-brane system,
we recover not only the tension and RR charge,
but also full effective action of the D$p$-brane exactly.
Our method works for general unstable D-brane systems
and provides a simple proof of D-brane descent/ascent relations
under the tachyon condensation.
In the lowest dimensional unstable D-brane system,
called K-matrix theory, D-branes are described in terms of
operator algebra.
We show the equivalence of the geometric and algebraic
descriptions of a D-brane world-volume manifold
using the equivalence between path integral and
operator formulation of the boundary quantum mechanics.
As a corollary, the Atiyah-Singer index theorem is naturally
obtained by looking at the coupling to RR-fields.
We also generalize the argument to type I string theory.

\end{titlepage}

\baselineskip=0.7cm

\tableofcontents

\section{Introduction}

A D-brane was first introduced in \cite{Pol et al} as
an extended object on which the end points of open strings
can be attached. Thanks to the duality between
open and closed strings,
it can also be seen as a source of
closed strings, which is described by a boundary state
\cite{CLNY,bdry}. On the other hand, D-branes are constructed
as non-commutative configurations in matrix models
\cite{BFSS}. More recently,
D-branes are realized as solitons in the gauge theory
with tachyon fields defined on higher dimensional
unstable D-brane systems \cite{Sen}.
The relation between the higher dimensional
unstable D-brane system and the lower dimensional
D-brane soliton is called D-brane descent relation.
{}From this construction, it has been revealed
that the D-brane charges are classified by K-theory
\cite{MiMo,WittenK,Ho}. Analogously,
it has been shown in \cite{Te,Kl,AsSuTe}
 that D-branes can also be constructed
as bound states of a lower dimensional unstable
D-brane system. We call this relation as D-brane ascent relation.
In \cite{AsSuTe}, construction
of D-branes from a matrix theory based on an unstable
D-instanton or D-particle system, which we call K-matrix theory,
is investigated in detail. In particular, it has been shown
that the classification of the D-branes is naturally given
by analytic K-homology,
which is consistent with the K-theory result mentioned above.
Note that in the latter two approaches, the tachyon
condensation
\footnote{For early works on tachyon condensation in string theory,
see \cite{Halpern}.}
 plays a crucial role in the construction
of D-branes.

In this paper, we will make the equivalence
among these different descriptions of D-branes
transparent. The basic idea is as follows.
We start with a boundary state representing
an unstable D-brane system, and turn on the boundary
interaction which represents the D$p$-brane solution.
Then, we show that the resulting boundary state is
nothing but the boundary state of the D$p$-brane.
This strategy has been taken in several works.
In \cite{Sen,Senbdry}, the boundary states with
non-trivial tachyon configurations are considered
in some special cases in which the boundary interaction
become exactly marginal operators.
In \cite{Is}, non-commutative D-branes
 are constructed by turning on
non-commutative configurations of scalar fields
in the boundary state representing infinitely many
 D-instantons. In \cite{AsSuTe}, this
strategy is used to construct D-brane
boundary states from the D-brane solutions in
 K-matrix theory.
One of the purposes of this paper is to extend the program
given in \cite{AsSuTe} to include fluctuations around
the D-brane solution. We will show that the fluctuations
around the D-brane solution correctly reproduce the fields
on the D-brane. We allow the fluctuations to be off-shell
so that we can apply the results to
the boundary string field theory (BSFT) \cite{WiBSFT}-\cite{TaTeUe}.
\footnote{Similar approach can also be found in \cite{deAl}.}
It immediately follows that the effective action for the
fluctuations around the D-brane solution
precisely reproduces the BSFT action for the D-brane.
Therefore, we recover not only the tension and
RR charge, but also full effective
action of the D$p$-brane exactly.
Since the boundary state contains all the information
about the D-brane, we think this provides a
formal proof of the D-brane descent/ascent relations.
Furthermore, we will develop a superfield formulation
of the boundary interaction, which was introduced
in \cite{AnTs}.
The derivation of the D-brane descent/ascent relations
is drastically simplified
by utilizing this superfield formulation.

Although our method works for general unstable D-brane systems,
we demonstrate it by mainly dealing with the lowest dimensional
unstable D-brane system, $i.e.$ the K-matrix theory.
K-matrix theory is the simplest set up
which can realize any D-brane configurations.
We consider a system with
 infinitely many non-BPS D-instantons
(for type IIA or type I) or D-instanton - anti D-instanton
pairs (for type IIB). Thus the matrix variables in the theory
are considered as linear operators acting on an infinite dimensional
Hilbert space. D-branes are represented algebraically
by using these operators. Actually, as it has been shown in
\cite{AsSuTe}, a D-brane is realized as (a limit of)
a spectral triple, which is introduced by Connes as
a realization of Riemannian manifold in non-commutative
geometry \cite{Connes}. This observation
makes it possible to naturally realize a D-brane
whose world-volume is a non-commutative space.
We will repeatedly see that
the equivalence of the geometric and algebraic
 descriptions of a D-brane
world-volume manifold
follows from the equivalence between
operator formalism and path integral formalism
of boundary quantum mechanics \cite{Is,AsSuTe}.
It is quite interesting that we can understand
the correspondence between operator algebra and geometry,
which is the starting point for
the non-commutative geometry,
using string theory.
Another application for the correspondence between
geometric and algebraic descriptions
is the Atiyah-Singer index theorem, which relates the index
of a Dirac operator with the Chern number of the gauge
bundle. We will give a direct physical interpretation for this
relation by examining the coupling to RR-fields.

We also generalize the argument to type I string theory.
The hidden real Clifford algebra structure of
the Chan-Paton Hilbert space for type I unstable
D-brane systems found in
\cite{AsSuTe2} plays a crucial role.
We will see that the Chern-Simons terms for type I D-branes
can be written down by using real superconnections.

The paper is organized as follows.
In section 2, we review the boundary states
with boundary interactions and the BSFT action
for unstable D-brane systems.
Section 3 is devoted to the construction of
D-branes in K-matrix theory.
We show that the fluctuations
around a D$p$-brane solution in K-matrix theory
correctly reproduce the fields on the D$p$-brane.
The D$p$-brane boundary states with the boundary
interactions are precisely reproduced from the boundary
state of unstable D-instanton system.
We generalize this argument to higher dimensional
systems in section 4.
In particular, the decent relations of D-branes
are shown in terms of the boundary states.
Section 5 deals with the generalization
to Type I string theory.
Finally we discuss further application
and some speculative discussions
in section 6.

\section{Boundary states and BSFT action}
\label{bdryrev}

In this section we review
the boundary states and the BSFT action
for unstable D-brane systems in type II string theory.
The D-brane boundary states
are obtained as linear combinations of
the boundary states defined by \cite{CLNY}
\footnote{We omit the ghost part which does not play any role
in this paper. See for example \cite{bdry}.}
\begin{eqnarray}
\ket{Bp;\pm}=
\int[dx^\alpha][d\psi^\alpha]
\ket{x^\alpha, x^i=0}\ket{\psi^\alpha, \psi^i=0;\pm},
\label{bpm}
\end{eqnarray}
where the superscript $\alpha$ and $i$ represent the direction
tangent and transverse to the D-brane, respectively.
The state $\ket{x}$ and $\ket{\psi}$ are the coherent states
satisfying
\begin{eqnarray}
X^\mu(\sigma)\ket{x}&=&x^\mu(\sigma)\ket{x},
\label{coherent}\\
\Psi_\pm^\mu(\sigma)\ket{\psi;\pm}&=&
\psi^\mu(\sigma)\ket{\psi;\pm},
\label{coherent2}
\end{eqnarray}
where\footnote{In this paper, we set $\alpha'=2$,
unless we recover explicit $\alpha'$ dependence.}
\begin{eqnarray}
X^\mu(\sigma)&=&\wh x_0^\mu+i\sum_{m\ne 0}
\left(
\frac{\alpha_m^\mu}{m}e^{-im\sigma}+\frac{\wt\alpha_m^\mu}{m}
e^{im\sigma}\right),\label{Xop}\\
\Psi^\mu_\pm(\sigma)&=&
\sum_{r}(\Psi^\mu_r e^{-ir\sigma}\pm i\wt\Psi^\mu_r e^{ir\sigma}),
\end{eqnarray}
are NS-R closed string operators at the boundary
of the world-sheet, which
act on the closed string Hilbert space, and
$x^\mu(\sigma)$ and $\psi^\mu(\sigma)$ are
bosonic and fermionic functions on $S^1$ ($0\le\sigma\le 2\pi$),
respectively.

The explicit form of the coherent states are given as follows.
\begin{eqnarray}
\ket{x}&=&
\exp\left\{\sum_{m=1}^\infty\left(-\half x_{-m} x_m-a^\dag_m\wt a^\dag_m
+a^\dag_m x_m+x_{-m} \wt a^\dag_m\right) \right\}\ket{x_0},
\label{ketx}\\
\ket{\psi;\pm}
&=&
\exp\left\{\sum_{r>0}\left(
-\half\psi_{-r}\psi_r \pm i\Psi^\dag_r\wt\Psi^\dag_r
+\Psi^\dag_r\psi_r\mp i\psi_{-r}\wt\Psi^\dag_r
\right)\right\}\ket{0},
\label{ketpsi}
\end{eqnarray}
where
\begin{eqnarray}
a_m^\mu=i\alpha_m^\mu/\sqrt{m},~~~
a_{-m}^\mu=a_m^{\mu\dag}=-i\alpha_{-m}^\mu/\sqrt{m},~~~(m>0),
\end{eqnarray}
and the $x_m$ and $\psi_r$ are the Fourier coefficients of
the eigen functions in (\ref{coherent}) and (\ref{coherent2});
\begin{eqnarray}
x^\mu(\sigma)=x_0^\mu+\sum_{m\ne 0}\frac{1}{\sqrt{|m|}}\,
x_m^\mu e^{-im\sigma},~~~
\psi^\mu(\sigma)
=\sum_{r}\psi_r^\mu e^{-ir\sigma}.
\end{eqnarray}
The state $\ket{x_0}$ in (\ref{ketx}) is the eigen state
of the zero mode $\wh x_0^\mu$ in (\ref{Xop}).
The equation (\ref{ketpsi}) is for the NSNS-sector.
For the RR-sector, we should replace $\ket{0}$ with
the eigen state  $\ket{\psi_0;\pm}$
for the fermion zero modes $\Psi_{0\pm}^\mu
=\Psi_{0}^\mu\pm i\wt\Psi_{0}^\mu$.
See \cite{AsSuTe} for our convention.

Fields on the D-brane world-volume can be turned on through
boundary interaction. Then,
the boundary states $\ket{Bp;\pm}$ are modified as
\begin{eqnarray}
\ket{Bp;\pm}_{S_b}&=&
e^{-S_b(X,\Psi_\pm)}\ket{Bp;\pm}
\label{BSb}
\\&=&
\int[dx^\alpha][d\psi^\alpha]
\,e^{-S_b(x,\psi)}
\ket{x^\alpha, x^i=0}\ket{\psi^\alpha, \psi^i=0;\pm}
\label{BSb2}
\end{eqnarray}
where $S_b$ is the boundary
action which represent the boundary interaction.
We often omit the subscript $\pm$ of the fermion $\Psi_\pm^\mu$
in the following.
The boundary interaction for the gauge fields on the
D-brane is well-known \cite{CLNY} and given as
a supersymmetric generalization of Wilson loop operator
\begin{eqnarray}
e^{-S_b(X,\Psi)}
=\TrP \exp\left\{-\int
d\sigma\left(A_\alpha(X)\dot X^\alpha
-\frac{1}{2}F_{\alpha\beta}(X)\Psi^\alpha\Psi^\beta
\right)\right\}.
\label{gauge}
\end{eqnarray}
Here, the gauge field $A_\alpha$ is taken to be
anti-hermitian matrix.

It is useful to write down this boundary state
using superfields.
Let us introduce superfields
$\bX^\mu(\wh\sigma)=X^\mu(\sigma)+i\theta\,\Psi^\mu(\sigma)$ and
$\bx^\mu(\wh\sigma)=x^\mu(\sigma)+i\theta\,\psi^\mu(\sigma)$,
where $\wh\sigma=(\sigma,\theta)$ is the boundary
supercoordinate.
The covariant derivative is defined as
$D=\del_\theta+\theta\del_\sigma$.
Then, the boundary interaction (\ref{gauge}) can be
written as \cite{AnTs}
\begin{eqnarray}
e^{-S_b(X,\Psi)}
=\TrwhP \exp\left(-\int
d\wh\sigma \,A_\alpha(\bX)D\bX^\alpha
\right),
\label{Sgauge}
\end{eqnarray}
where $d\wh\sigma=d\sigma d\theta$.
Here $\wh{\rm P}$ denotes the supersymmetric
path ordered product, which is defined as
\begin{eqnarray}
&&\wh{\rm P}\,\exp\left(\int d\hat\sigma \,\bM(\hat\sigma)\right)\nn\\
&=&
\sum_{n=0}^\infty(-1)^{\frac{n(n-1)}{2}}
\int d\wh\sigma_1\cdots d\wh\sigma_n
\Theta(\wh\sigma_{12})\Theta(\wh\sigma_{23})\cdots
\Theta(\wh\sigma_{n-1\,n})
\,\bM(\wh\sigma_1)\cdots\bM(\wh\sigma_n),\nn\\
\label{SP}
\end{eqnarray}
where $\wh\sigma_{ab}=\sigma_a-\sigma_b-\theta_a\theta_b$
and $\Theta$ is a step function.
If we perform the $d\theta$ integral in (\ref{SP}),
we obtain the ordinary path ordered product as
\begin{eqnarray}
\wh{\rm P}\,e^{\int d\hat\sigma\bM(\hat\sigma)}
={\rm P}\,e^{\int d\sigma (M_1-M_0^2)(\sigma)},
\label{SP2}
\end{eqnarray}
where we write $\bM(\wh\sigma)=M_0(\sigma)+\theta\,
M_1(\sigma)$.
Note that $M_0^2$ term in (\ref{SP2}) comes from
the contact $\delta$-function part of the expansion
$\Theta(\wh\sigma_{ab})
=\Theta(\sigma_a-\sigma_b)-\theta_a\theta_b\delta(\sigma_a-\sigma_b)$.
Using this formula (\ref{SP2}), it is easy to recover
(\ref{gauge}) from (\ref{Sgauge}).

The boundary interactions
for space-time filling unstable D-brane
systems (non-BPS D9-branes in type IIA and
D9-\AD9 system in type IIB),
which include gauge fields and tachyon fields,
are obtained in \cite{HaKuMa,KuMaMo2,KrLa,TaTeUe}.
It is straightforward to generalize their argument
to lower dimensional D-brane systems as described below.

First, we introduce a matrix consists of a condensate of
the open string vertices in the superfield notation
\begin{eqnarray}
\bM=\mat{-A_\alpha(\bX) D\bX^\alpha
-i\Phi^i(\bX)\bP_i,T(\bX),T(\bX)^\dag,
-\wt A_\alpha(\bX) D\bX^\alpha-i\wt\Phi^i(\bX)\bP_i},
\label{superM}
\end{eqnarray}
where $\bP_i(\wh\sigma)=\theta\,P_i(\sigma)+i\Pi_i(\sigma)$.
Here $P_i(\sigma)$ and $\Pi_i(\sigma)$ are the conjugate
momenta of $X^i(\sigma)$ and $\Psi^i(\sigma)$, respectively.
$A_\alpha$, $\wt A_\alpha$, $\Phi^i$, $\wt\Phi^i$ and $T$
are the fields on the D-brane.
\footnote{We do not consider massive modes and fermions
in this paper.}
These fields are independent for D$p$-\AD{$p$}-brane system.
Namely, $A_\alpha$ and $\Phi^i$ are the gauge field and scalar fields
on the D$p$-branes,
$\wt A_\mu$ and $\wt\Phi^i$ are those on the
\AD{$p$}-branes and $T$ is the tachyon field which
is created by the open string stretched between
the D$p$-branes and the \AD{$p$}-branes.
For non-BPS D$p$-brane, we have constraints
 $A_\alpha=\wt A_\alpha$,
$\Phi^i=\wt\Phi^i$ and $T^\dag=T$.

Let us consider $N$ pairs of D$p$-brane and \AD$p$-brane.
We can decompose the matrix (\ref{superM}) using Pauli matrices
 $\sigma_1$ and $\sigma_2$ as
\begin{eqnarray}
\bM&=&
-(A_\alpha^{+}D\bX^\alpha+i\Phi^i_{+}\bP_i)\otimes 1_2
-(A_\alpha^{-}D\bX^\alpha+i\Phi^i_{-}\bP_i)
\otimes i\sigma_2\sigma_1\nonumber\\
&&~~+T_1\otimes\sigma_1+T_2\otimes \sigma_2,
\label{bM}
\end{eqnarray}
where $A_\alpha^{\pm}=\half(A_\alpha\pm \wt A_\alpha)$,
 $\Phi^i_{\pm}=\half(\Phi^i\pm \wt\Phi^i)$,
$T_1=\half(T+T^\dag)$ and $T_2=\frac{i}{2}(T- T^\dag)$.
Then, the boundary interaction can be written as
\begin{eqnarray}
e^{-S_b}&=&\int[d\bG^1][d\bG^2]
\,\Tr\wh{\rm P}\,\exp\left\{
\int d\wh\sigma \left(
\frac{1}{4}\bG^1 D\bG^1+\frac{1}{4}\bG^2 D\bG^2\right.\right.
\nonumber\\
&&~~~~~
-(A_\alpha^{+}D\bX^\alpha+i\Phi^i_{+}\bP_i)
-(A_\alpha^{-}D\bX^\alpha+i\Phi^i_{-}\bP_i)\,i\bG^2\bG^1
\nonumber\\
&&~~~~~~~~\left.\left.+T_1\bG^1+T_2\bG^2
\frac{}{}\right)\right\},
\label{bdrybM}
\end{eqnarray}
where
 $\bG^I(\wh\sigma)=\eta^I(\sigma)+\theta\,F^I(\sigma)$ ($I=1,2$)
are real fermionic superfields.
Here the trace $\Tr$ is taken over remaining
$N$ Chan-Paton indices.
The basic idea to obtain (\ref{bdrybM}) from the matrix
(\ref{bM}) is to replace the Pauli matrices $\sigma^I$ with
fields $\eta^I$ which satisfy the same algebra as $\sigma^I$, $i.e.$
\begin{eqnarray}
\{\wh\eta^I,\wh\eta^I\}=2\delta^{IJ},
\end{eqnarray}
in the operator formulation and combine them with their
superpartner fields $F^I$ in the superfields $\bG^I$.

This prescription can easily be generalized to the case
in which the matrix $\bM$ is expanded by $SO(2m)$
gamma matrices $\Gamma^I=\left({~~~\gamma^I\atop\gamma^{I\dag}~~~}\right)$
 ($I=1,\dots,2m$)
as
\begin{eqnarray}
\bM=\sum_{k=0}^{2m} \bM^{I_1\cdots I_k}
\otimes\Gamma^{I_1\cdots I_k},
\label{gammaexp}
\end{eqnarray}
where
 $\Gamma^{I_1\cdots I_k}$ denote the
skew-symmetric products of the gamma matrices.
In this case, the boundary interaction becomes \cite{KrLa}
\begin{eqnarray}
e^{-S_b}=\int[d\bG^I]\,\Tr\wh{\rm P}\,
\exp\left\{
\int d\wh\sigma \left(
\frac{1}{4}\bG^I D\bG^I
+\sum_{k=0}^{2m} \bM^{I_1\cdots I_k}
\bG^{I_1}\cdots\bG^{I_k}
\right)\right\}.
\label{super}
\end{eqnarray}
Here, the path ordered product and the trace $\TrP$
is needed when $\bM^{I_1\dots I_k}$
are still matrices.
Note that $F^I$ are auxiliary fields which can be integrated out.
After performing $\theta$ integral and integrating out the
auxiliary fields $F^I$, we obtain \cite{KrLa}
\begin{eqnarray}
e^{-S_b}=\int[d\eta^I]\TrP\exp\left\{
\int d\sigma\left(
\frac{1}{4}\dot\eta^I\eta^I+
\sum_{k=0}^{2m}
M^{I_1\cdots I_k}\eta^{I_1}\cdots\eta^{I_k}
\right)
\right\},
\label{bint}
\end{eqnarray}
where
\begin{eqnarray}
M&=&\wh\cA+\wh\cF,
\label{MAF}\\
&=&\sum_{k=0}^{2m} M^{I_1\cdots I_k}
\otimes\Gamma^{I_1\cdots I_k},
\label{MAFexp}
\end{eqnarray}
\begin{eqnarray}
\wh\cA&=&
\mat{\cA,,,\wt\cA},\\
\wh\cF&=&\mat{-TT^\dag+\cF,\cD T,\cD T^\dag,-T^\dag T+\wt\cF},
\label{whF}
\end{eqnarray}
\begin{eqnarray}
\cA&=&-A_\alpha \dot X^\alpha-i\Phi^i P_i
\label{cA}\\
\wt\cA&=&-\wt A_\alpha \dot X^\alpha-i\wt\Phi^i P_i\\
\cF&=&
\frac{1}{2}F_{\alpha\beta}\Psi^\alpha\Psi^\beta
+i(\del_\alpha\Phi^i+[A_\alpha,\Phi^i])\Psi^\alpha\Pi_i
-\half[\Phi^i,\Phi^j]\Pi_i\Pi_j
\label{cF}\\
\wt\cF&=&
\frac{1}{2}\wt F_{\alpha\beta}\Psi^\alpha\Psi^\beta
+i(\del_\alpha\wt\Phi^i+[\wt A_\alpha,\wt\Phi^i])\Psi^\alpha\Pi_i
-\half[\wt\Phi^i,\wt\Phi^j]\Pi_i\Pi_j\\
\cD T&=&i(\del_\alpha T+A_\alpha T-T\wt A_\alpha)\Psi^\alpha
-(\Phi^i T-T\wt\Phi^i)\Pi_i
\label{cDT}\\
\cD T^\dag&=&i(\del_\alpha T^\dag+\wt A_\alpha T^\dag-T^\dag A_\alpha)
\Psi^\alpha-(\wt\Phi^i T^\dag-T^\dag\Phi^i)\Pi_i.
\end{eqnarray}
It is also useful to note that $\wh\cF$ can be written as
\begin{eqnarray}
\wh\cF=-Z^2,~~~
Z=\mat{-i\cD,T,T^\dag,-i\wt\cD},
\end{eqnarray}
where $\cD=\Psi^\alpha(\del_\alpha+A_\alpha)+i\Pi_i\Phi^i$
and  $\wt\cD=\Psi^\alpha(\del_\alpha+\wt A_\alpha)+i\Pi_i\wt\Phi^i$.

If we further perform the integral over $\eta^I$,
the boundary interaction (\ref{bint}) becomes
\begin{eqnarray}
e^{-S_b}&=&
\left\{
\begin{array}{cl}
\kappa\TrP\,e^{\int\! d\sigma M(\sigma)}&(\mbox{NS-NS sector}),\\
\Str {\rm P}\,e^{\int\! d\sigma M(\sigma)}&(\mbox{R-R sector}),
\end{array}
\right.
\label{supertrace}
\end{eqnarray}
where $M$ is given as (\ref{MAF}).
The normalizaton constant $\kappa$ is
$\kappa=1$ for D$p$-\AD$p$ systems and $\kappa=1/\sqrt{2}$
for non-BPS D-branes.
See Appendix \ref{Oppath} and \ref{Normalization}
for the derivation.
Note that, in computing the products of the matrix $M$,
$\Gamma^I$ in the gamma matrix expansion
(\ref{MAFexp}) are treated as fermionic gamma
matrices, which anti-commute with $\Psi^\mu$ and $\Pi_\mu$,
since $\eta^I$ are fermionic. (See \cite{KrLa}.)
This formula (\ref{supertrace}) reduces to (\ref{gauge}) when
we only turn on the gauge field $A_\alpha$.
Here $\Str$ denote the supertrace defined as
$\Str(\cdots)=\kappa\Tr(\Gamma\cdots)$, where
$\Gamma=(-i)^{n/2}\prod_{I=1}^n\Gamma^I$
 is the skew-symmetric product
of all the gamma matrices $\Gamma^I$ ($I=1,\dots,n$)
 which take place
in the gamma matrix expansion.
\footnote{Note that the normalization of the supertrace
defined here is slightly different from the usual one
when $n$ is odd, $i.e.$ for the non-BPS D-branes.
We adopt this normalization in order to treat
the D$p$-\AD$p$ system and non-BPS D$p$-branes in
the same notation.}
The difference between NS-NS sector and R-R sector
comes from the boundary condition of the boundary
fermion fields.
In the R-R sector, $\eta^I(\sigma)$ are periodic
with respect to $\sigma$ and there is a zero mode for
each of them. Therefore,
we have to saturate the zero modes in the path integral
(\ref{bint}), implying the supertrace rather than
the usual trace.

For the D$p$-\AD$p$ system, we can see from
 (\ref{MAF})--(\ref{whF}) that $e^{\int d\sigma M}$ is
of the form
\begin{eqnarray}
\mat{\alpha,\beta,\gamma,\delta},
\end{eqnarray}
where $\alpha$, $\beta$, $\gamma$
 and $\delta$ are the components related to
D$p$-D$p$ string, D$p$-\AD$p$ string,
 \AD$p$-D$p$ string and \AD$p$-\AD$p$ string, respectively.
This is also the case for non-BPS D-branes,
if we impose the constraints $\alpha=\delta$ and $\beta=\gamma$.
Thus this matrix can be expanded
with respect to $\sigma_1$ and $\sigma_2$
for the D$p$-\AD$p$ system, just as we did in (\ref{bM}),
while only $\sigma_1$ is needed for the non-BPS D-branes.
Therefore, we have
\begin{eqnarray}
\Str\mat{\alpha,\beta,\gamma,\delta}
&=&\left\{
\begin{array}{cl}
\Tr\left[\,(-i)\,\sigma_1\sigma_2
\left({\alpha\,\beta\atop\gamma\,\delta}\right)
\right],
&(\mbox{D$p$-\AD$p$ system}),\\
\frac{1}{\sqrt{2}}\Tr\left[\,(-i)^{1/2}\sigma_1
\left({\alpha\,\beta\atop\beta\,\alpha}\right)
\right],
&(\mbox{non-BPS D$p$-branes}),
\end{array}
\right.\\
&=&\left\{
\begin{array}{cl}
\tr\alpha-\tr\delta,&(\mbox{D$p$-\AD$p$ system}),\\
(-2i)^{1/2}\tr\beta,&(\mbox{non-BPS D$p$-branes}).
\end{array}
\right.
\label{str}
\end{eqnarray}
The formula (\ref{supertrace}) together with
(\ref{str}) can also be used in the case
with different numbers of D$p$-branes and \AD$p$-branes.
In particular, we can consider BPS D$p$-branes by
setting the size of the matrices
$\beta$, $\gamma$ and $\delta$ to be zero, which
again reproduces (\ref{gauge}).

Note also that (\ref{MAF}) is given as $M=M_1-M_0^2$,
where $\bM=M_0+\theta M_1$ is the matrix in (\ref{superM}),
and hence
the boundary interaction (\ref{supertrace}) can also be written
in superfields as
\begin{eqnarray}
e^{-S_b}&=&
\left\{
\begin{array}{cl}
\kappa\TrwhP\,e^{\int\! d\hat\sigma
 \bM(\hat\sigma)}&(\mbox{NS-NS sector}),\\
\Str\wh{\rm P}\,e^{\int\! d\hat\sigma \bM(\hat\sigma)}
&(\mbox{R-R sector}),
\end{array}
\right.
\label{supertrP}
\end{eqnarray}
using the relation (\ref{SP2}).
See Appendix \ref{Oppath}
for a direct derivation of the equivalence between (\ref{super})
and (\ref{supertrP}).
To sum up, we can use any one of
 (\ref{super}), (\ref{bint}),
(\ref{supertrace}) and (\ref{supertrP}) as the boundary
interaction.

The boundary state for a D$p$-brane is given by a
linear combination of $\ket{Bp;\pm}$ with the boundary
interaction such that
\begin{eqnarray}
\ket{Dp}_{S_b}&=&
P\wt P_+ \,e^{-S_b}\ket{Bp;+}_{\rm NS}+
P\wt P_\pm \,e^{-S_b}\ket{Bp;+}_{\rm RR},
\label{GSO}
\end{eqnarray}
where $P=\half(1+(-1)^F)$ and  $\wt P_\pm=\half(1\pm(-1)^{\wt F})$
are GSO projection operators. The subscript of
$\wt P_\pm$ in the right hand side of (\ref{GSO})
is $+$ or $-$ for type IIB or type IIA string theory, respectively.

The boundary string field theory (BSFT) action for the D$p$-brane
in superstring theory is
given in \cite{KuMaMo2,KrLa,TaTeUe} as a disk partition function
with the boundary interaction, which can be
written as
\begin{eqnarray}
S(T,A_\alpha,\Phi^i,\cdots)=\frac{2\pi}{g_s}
\bra{0} e^{-S_b} \ket{Bp;+}_{\rm NS},
\label{BSFT}
\end{eqnarray}
using the NS-NS sector boundary state with the boundary interaction
(\ref{BSb}).
\footnote{
The normalization factor is
determined in Appendix \ref{Normalization}.}
Similarly the CS-term of the system is given by
\begin{eqnarray}
S_{CS}(C,T,A_\alpha,\Phi^i,\cdots)
=\bra{C} e^{-S_b} \ket{Bp;+}_{\rm RR},
\label{CS}
\end{eqnarray}
where $\bra{C}$ is the state corresponding to the R-R fields.

There are many attempts to calculate the action (\ref{BSFT})
or the CS-term (\ref{CS}) explicitly.
When the tachyon is absent, up to four
derivative corrections to the DBI action is calculated
in \cite{Wy} using the boundary state with
the boundary interaction (\ref{Sgauge}).
The CS-term is examined using the expression (\ref{CS})
in \cite{DiVeFrLeLi} and its derivative corrections
are also calculated in \cite{Wy}.
For the unstable D-brane systems,
the action has been obtained explicitly by now only
for constant $\del_\mu T$ and $F_{\mu \nu}$ \cite{TeUe}
or up to the order $(\alpha')^2$ \cite{TaTeUe}.
However, we stress here that
the boundary action of the BSFT action
is renormalizable if we consider the tachyon and gauge fields only.
Thus the BSFT action is well-defined for those fields
and we can in principle calculate all the
higher derivative corrections using the expression
(\ref{BSFT}).

\section{D-branes in K-matrix theory}
\label{DinKmat}
In this section, we consider the construction of D-branes
in a matrix theory based on an unstable D-brane system,
which we call K-matrix theory. There are many choices of
the unstable D-brane system. The lowest dimensional ones
are non-BPS D-instanton system in type IIA string theory
or D-instanton - anti D-instanton system in type IIB
string theory.
If we want to avoid the formal Euclidean rotation
to obtain the D-instantons, we could start with
D$0$-\AD$0$ system in type IIA string theory or
non-BPS D$0$-brane system in type IIB string theory.

In this section, we will mainly use
the matrix theory based on the non-BPS D-instantons
in type IIA string theory for simplicity.
The generalization to other unstable D-brane systems
is straightforward.

\subsection{K-matrix theory and D-brane solutions}

Let us consider
 the matrix theory based on
non-BPS D-instanton system in type IIA string theory.
The field contents and the action is obtained as
the dimensional reduction of non-BPS D9-brane system.
The $N$ non-BPS D9-brane system is a ten dimensional gauge theory
with $U(N)$ gauge symmetry with a tachyon field.
In this paper, we only consider low-lying bosonic
fields, that is, the tachyon field $T$ and the gauge field $A_\mu$,
both of which transform as the adjoint representation under
the gauge group.
The dimensional reduction of these fields gives
the tachyon $T$ and ten scalar fields $\Phi^\mu$ ($\mu=0,1,\dots,9$).
They are $N\times N$ hermitian matrices. We take $N$ to be infinity
and regard these matrices as operators acting on an infinite
dimensional Hilbert space.
The BSFT action (\ref{BSFT}) of the system is
\begin{eqnarray}
S(\Phi^\mu,T)=\frac{2\pi}{g_s}\bra{0}e^{-S_b(\Phi^\mu,T)}
\ket{B(-1);+}_{\rm NS},
\end{eqnarray}
where the boundary interaction is given by (\ref{super})
or (\ref{supertrP}) with
\begin{eqnarray}
\bM=\mat{-i\Phi^\mu\bP_\mu,T,T,-i\Phi^\mu\bP_\mu},
\end{eqnarray}
or,  (\ref{bint}) or (\ref{supertrace}) with
\begin{eqnarray}
M= \mat{-i\Phi^\mu P_\mu-T^2-\half{[\Phi^\mu,\Phi^\nu]}
\,\Pi_\mu\Pi_\nu, -{[\Phi^\mu,T]}\,\Pi_\mu,
-{[\Phi^\mu, T]}\,\Pi_\mu,
-i\Phi^\mu P_\mu-T^2-\half{[\Phi^\mu,\Phi^\nu]}\,\Pi_\mu\Pi_\nu}\,.
\label{M}
\end{eqnarray}

As argued in \cite{AsSuTe}, we can construct any
D-brane configuration expected in type IIA string theory
using this matrix theory.
In fact, it has been shown in \cite{AsSuTe}
that the D-brane configurations are given as
(a limit of) spectral triples, which are analytic
description of Riemannian manifold \cite{Connes}, and they are
 classified by K-homology, which is consistent with
the K-theory classification of D-brane charges
\cite{WittenK,Ho}.

More explicitly, a solution representing
a D$p$-brane extended along $x^0,\dots,x^p$-directions
is given by
\begin{eqnarray}
T&=&u\,\sum_{\alpha=0}^p
\wh{p}_\alpha \otimes \gamma^\alpha
\label{sol1-1}
\\
\Phi^{\alpha} &=& \wh{x}^{\alpha} \otimes {\rm 1}
\quad
(\alpha=0, \dots, p)\,,
\qquad
\Phi^i=0
\quad
(i=p+1, \dots, 9)\,,
\label{sol1-2}
\end{eqnarray}
where $\wh x^\alpha$ and $\wh p_\alpha$ are operators
on a Hilbert space $\cH$ satisfying
\begin{eqnarray}
[\wh x^\alpha,\wh p_\beta]=i\delta^\alpha_\beta,
\label{ccr}
\end{eqnarray}
and $\gamma^\alpha$ are gamma matrices represented as
hermitian matrices. $u$ is a real parameter and
this configuration becomes an exact solution in the limit
$u\ra \infty$ \cite{Te}.
Since the eigen values of $\Phi^\mu$ represent the position
of the non-BPS D-instantons, we can see from (\ref{sol1-2})
that this configuration represent a $p+1$ dimensional
object, which is interpreted as the D$p$-brane world volume.

The fields on the D$p$-brane are given by
the fluctuation around this solution (\ref{sol1-1}, \ref{sol1-2}).
Our goal is to consider all the relevant fluctuations
around this solution and identify them with the fields
on the D$p$-branes and then
show that we can exactly reproduce
the effective action of the D$p$-brane (\ref{BSFT}).
For this purpose, it suffices to show
that the boundary state of the D$p$-brane $e^{-S_b}\ket{Bp,\pm}$
is reproduced by inserting the D$p$-brane solution and
its fluctuations into the boundary state
of the non-BPS D-instantons $e^{-S_b(\Phi^\mu,T)}\ket{B(-1),\pm}$.
Actually, since the boundary state carries all the information
about the D-brane, this explicitly shows that
the D$p$-brane is constructed from the non-BPS D-instantons.
We emphasize that we recover not only the tension,
RR charge or tachyon mass etc, but also full effective
action of the D$p$-brane
as well as its coupling to every closed string mode exactly.

We start by introducing the fluctuations one by one
in section \ref{gaugefield}--\ref{tachyonfield} to
identify them with the fields on the D$p$-brane.
We shall analyze more general fluctuations
in section \ref{general} after developing a formulation
using superfield. The analysis
will become surprisingly simple using this superfield method.
Hasty readers could skip
section \ref{gaugefield}--\ref{tachyonfield} and
proceed to section \ref{general}.
The following sections, section \ref{cs-term} and
section \ref{NCGT}, are devoted to some applications
of our argument.

\subsection{The gauge field}
\label{gaugefield}

Let us first consider the fluctuation which
corresponds to the gauge field on the D$p$-brane.
A solution representing $N$ D$p$-branes can be obtained
by piling $N$ copies of a single D$p$-brane solution
(\ref{sol1-1}, \ref{sol1-2}) as
\begin{eqnarray}
T&=&u\,\sum_{\alpha=0}^p
\wh{p}_\alpha \otimes 1_N\otimes \gamma^\alpha
\label{sol2-1}
\\
\Phi^{\alpha} &=& \wh{x}^{\alpha}\otimes 1_N \otimes {\rm 1}
\quad
(\alpha=0, \dots, p)\,,
\qquad
\Phi^i=0
\quad
(i=p+1, \dots, 9)\,,
\label{sol2-2}
\end{eqnarray}
Here,  $T$ and $\Phi^\mu$ are operators acting on a Hilbert space
$\cH\otimes \C^N\otimes S$, where $S$ is the spinor space on which
the gamma matrices $\gamma^\alpha$ are represented.
The gauge symmetry of the system is given by the
transformation
\begin{eqnarray}
T\ra UTU^{-1},~~~\Phi^\mu\ra U\Phi^\mu U^{-1}.
\label{gaugetr}
\end{eqnarray}
with a unitary operator $U$.
The global gauge symmetry of the D$p$-brane is the
part of the unitary symmetry which is
unbroken in the configuration (\ref{sol2-1}) and (\ref{sol2-2}),
namely, the $U(N)$ subgroup of the form $U=1\otimes h\otimes 1$
($h\in U(N)$). The local gauge symmetry of the D$p$-brane
is obtained by allowing the $\wh x$ dependence in the
unitary transformation as $U=h(\wh x)\otimes 1$,
where $h(x)$ is a $U(N)$ valued function.
 The gauge field corresponding to this is introduced in the
tachyon operator as
\begin{eqnarray}
T&=&u\,\sum_{\alpha=0}^p
(\wh{p}_\alpha-iA_\alpha(\wh{x})) \otimes \gamma^\alpha
\label{AinT}
\end{eqnarray}
in order to maintain the covariance under the local gauge
transformation \cite{AsSuTe}. Here we abbreviated
$\wh p_\alpha\otimes 1_N$ as $\wh p_\alpha$.

Inserting (\ref{AinT}) and (\ref{sol2-2}) into (\ref{M}), we obtain
\begin{eqnarray}
M &=& \mat{-i\wh x^\alpha P_\alpha-u^2(\wh p_\alpha-iA_\alpha)^2
+\frac{u^2}{2} F_{\alpha\beta}\gamma^{\alpha\beta},
-iu\gamma^\alpha\Pi_\alpha,-iu\gamma^\alpha\Pi_\alpha,
-i\wh x^\alpha
P_\alpha-u^2(\wh p_\alpha-iA_\alpha)^2
+\frac{u^2}{2} F_{\alpha\beta}\gamma^{\alpha\beta}}\nn\\ \\
&=&-\left(i\wh x^\alpha P_\alpha+u^2(\wh p_\alpha-iA_\alpha)^2\right) I
-iu\Pi_\alpha\Gamma^\alpha
+\frac{u^2}{2} F_{\alpha\beta}\Gamma^{\alpha\beta},
\end{eqnarray}
where we have defined
$\Gamma^\alpha={~~~\gamma^\alpha\choose \gamma^\alpha~~~}$.
Here we adopt (\ref{bint}) for the boundary interaction.
Then, we obtain
\begin{eqnarray}
&&e^{-S_b(\Phi^\mu,T)}\\
&=& \int[d\eta^\alpha]\, \Tr{\rm P} \exp\left\{
\int d\sigma\left( \frac{1}{4}\dot\eta^\alpha\eta^\alpha- i\wh
x^\alpha P_\alpha- u^2(\wh p_\alpha-iA_\alpha)^2
-iu\Pi_\alpha\eta^\alpha
+\frac{u^2}{2}F_{\alpha\beta}\eta^\alpha\eta^\beta
\right)\right\}\nn\\
\\
&=& \int[d\eta^\alpha]\, e^{ \int d\sigma\left(
\frac{1}{4}\eta^\alpha\dot\eta^\alpha-iu\Pi_\alpha\eta^\alpha\right)
} \Tr{\rm P}\, e^{-i\int\! d\sigma H(\hat p,\hat x)}\,,
\label{Bint2}
\end{eqnarray}
where $\Tr$ denotes the trace over $\cH\otimes\C^N$
and we defined
$i H(\wh p,\wh x)=u^2(\wh p_\alpha-iA_\alpha(\wh x))^2
+i\wh x^\alpha P_\alpha
-\frac{u^2}{2}F_{\alpha\beta}(\wh x)\eta^\alpha\eta^\beta$.
Since the last term in (\ref{Bint2}) is usual quantum mechanical
partition function with Hamiltonian $H(\wh p,\wh x)$,
we can rewrite it in terms
of the path integral formulation using the standard formula
\begin{equation}
\Tr{\rm P}\, e^{-i\int\! d\sigma H(\hat p,\hat x)}
=\int[dx]\trP\, e^{i\int\! d\sigma L(\dot x,x)}\,,
\label{key}
\end{equation}
where $\tr$ in the right hand side denotes
the trace over $\C^N$, $i.e.$, with respect to the gauge indices
of the resulting D$p$-branes.
The Lagrangian $L(\dot x,x)$ is obtained by the Legendre transformation
of the Hamiltonian $H(\wh p,\wh x)$
\begin{eqnarray}
iL(\dot x,x)=-\frac{\dot x_{\alpha}^2}{4u^2}-A_\alpha(x)\dot x^\alpha
-i x^\alpha P_\alpha
+\frac{u^2}{2}F_{\alpha\beta}(x)\eta^\alpha\eta^\beta.
\end{eqnarray}
Thus we obtain
\begin{eqnarray}
&&e^{-S_b(\Phi^\mu,T)}\\
&=& \int[dx^\alpha][d\eta^\alpha]\trP\, \exp\left\{ \int
d\sigma\left( \frac{1}{4}\dot\eta^\alpha\eta^\alpha
-\frac{\dot x_\alpha^2}{4u^2}
\right.\right.\nn
\\
&&\hspace{16ex}\left.\left.
-A_\alpha(x)\dot x^\alpha
-ix^\alpha P_\alpha
+\frac{u^2}{2}F_{\alpha\beta}(x)\eta^\alpha\eta^\beta
-iu\Pi_\alpha\eta^\alpha
\right)\right\}
\\
&=& \int[dx^\alpha][d\psi^\alpha]\trP\, \exp\left\{\int
d\sigma\left( -\frac{1}{4u^2}(\dot
x_\alpha^2+\psi^\alpha\dot\psi^\alpha)\right.\right.\nn\\
&&\hspace{16ex}\left.\left.
-A_\alpha(x)\dot x^\alpha
+\frac{1}{2}F_{\alpha\beta}(x)\psi^\alpha\psi^\beta
-ix^\alpha P_\alpha-i\Pi_\alpha\psi^\alpha \right)\right\},
\label{spinning}
\end{eqnarray}
where $\psi^\alpha=u\eta^\alpha$.
\footnote{
We can show $[d\eta]=[ud\psi^\alpha]=[d\psi^\alpha]$
using the zeta-function regularization \cite{AsSuTe}.
}
Taking the limit $u\ra\infty$, we obtain
\footnote{To be more precise, we should regularize the path integral
before taking the $u\ra\infty$ limit. The following
procedure can be justified by using zeta-function regularization
\cite{AsSuTe}.
}
\begin{eqnarray}
&&e^{-S_b(\Phi^\mu,T)}\nn\\
&\ra& \int[dx^\alpha][d\psi^\alpha]\trP\, \exp\left\{ \int
d\sigma\left(-A_\alpha(x)\dot x^\alpha
+\frac{1}{2}F_{\alpha\beta}(x)\psi^\alpha\psi^\beta
-ix^\alpha P_\alpha-i\Pi_\alpha\psi^\alpha \right)\right\},\nn\\
\end{eqnarray}
and the boundary state will become
\begin{eqnarray}
&&e^{-S_b(\Phi^\mu,T)}\ket{B(-1);\pm}\\
&\ra& \int[dx^\alpha][d\psi^\alpha]\trP\, \exp\left\{ \int\!
d\sigma\left(-A_\alpha(x)\dot x^\alpha
+\frac{1}{2}F_{\alpha\beta}(x)\psi^\alpha\psi^\beta
\right.\right.\nn\\
&&\left.\left.\hspace{25ex}
-ix^\alpha P_\alpha\!
-i\Pi_\alpha\psi^\alpha
\frac{}{}
\right)\right\}
\ket{x^\mu=0}\ket{\psi^\mu=0;\pm}
\\
&=& \int[dx^\alpha][d\psi^\alpha]\trP\, \exp\left\{\int
d\sigma\left(-A_\alpha(x)\dot x^\alpha
+\frac{1}{2}F_{\alpha\beta}(x)\psi^\alpha\psi^\beta
\right)\right\}\nn\\
&&\hspace{25ex}\times
\ket{x^\alpha, x^i=0}\ket{\psi^\alpha,\psi^i=0;\pm}
\\
&=&e^{-S_b(A_\alpha)}\ket{Bp;\pm},
\label{uinftybdry}
\end{eqnarray}
where
\begin{eqnarray}
e^{-S_b(A_\alpha)}
=\trP \exp\left\{\int
d\sigma\left(-A_\alpha(X)\dot X^\alpha
+\frac{1}{2}F_{\alpha\beta}(X)\Psi^\alpha\Psi^\beta
\right)\right\}.
\end{eqnarray}
(\ref{uinftybdry}) is nothing but the boundary state for D$p$-brane
(\ref{BSb})
with the boundary interaction (\ref{gauge}).

Note that (\ref{spinning}) is the partition function for
the supersymmetric quantum mechanics of a spinning particle
in an electro-magnetic background (with source terms),
although the supersymmetry is broken by the boundary condition
for the NS-NS sector.
What we have done here is to rewrite
the trace over the Chan-Paton Hilbert space $\cH$
using the path integral formulation of the
boundary quantum mechanics.
This is the essence for reproducing
the D$p$-brane boundary state from
that described as a bound state of
infinitely many non-BPS D-instantons.
We can also see that the solution
part (\ref{sol2-1}), (\ref{sol2-2}) reproduces
the base space $\ket{Bp;\pm}$, while the fluctuation part
in (\ref{AinT}) contributes only to the boundary interaction
$e^{-S_b}$ for the D$p$-brane.

\subsection{The scalar fields}
\label{scalarfield}

The massless scalar fields on the D$p$-brane represent the position
of the D$p$-brane world-volume in the transverse direction.
The D$p$-brane is now constructed by non-BPS D-instantons
whose position is represented by eigen values of
the operators $\Phi^\mu$.
Therefore, the fluctuations corresponding to
the scalar fields on the
D$p$-brane is obtained by considering the fluctuation
\begin{eqnarray}
\Phi^i=\phi^i(\wh x),~~~~(i=p+1,\dots,9)
\label{Phifluc0}
\end{eqnarray}
around the solution $\Phi^i=0$ in (\ref{sol2-2}).
Here we abbreviated $\phi^i(\wh x)\otimes 1$
on $(\cH\otimes\C^N)\otimes S$ as $\phi^i(\wh x)$.
Then, (\ref{M}) becomes
\begin{eqnarray}
M
&=&-\left(i\wh x^\alpha P_\alpha+i\phi^i P_i+\half[\phi^i,\phi^j]
\Pi_i\Pi_j
+u^2(\wh p_\alpha-iA_\alpha)^2\right) I\nn\\
&&~~~
-iu\left(\Pi_\alpha+ D_\alpha\phi^i \Pi_i\right)\Gamma^\alpha
+\frac{u^2}{2}F_{\alpha\beta}\Gamma^{\alpha\beta},
\label{M2}
\end{eqnarray}
Repeating the analysis in the previous subsection,
we obtain
\begin{eqnarray}
e^{-S_b(\Phi^\mu,T)}\ket{B(-1);\pm}
\ra e^{-S_b(A_\alpha,\phi^i)}\ket{Bp;\pm}
\end{eqnarray}
in the $u\ra\infty$ limit. Here
\begin{eqnarray}
e^{-S_b(A_\alpha,\Phi^i)}
=\trP \exp\left\{
\int d\sigma\left(
\cA+\cF
\right)
\right\},
\end{eqnarray}
where $\cA$ and $\cF$ are defined in (\ref{cA}) and (\ref{cF}).
This reproduces the correct $\phi^i(x)$ dependence
in the boundary interaction for D$p$-brane
reviewed in section \ref{bdryrev}.

\subsection{The tachyon field}
\label{tachyonfield}

When $p$ is odd, the solution
(\ref{sol2-1}, \ref{sol2-2}) represents $N$ non-BPS D$p$-branes
in type IIA string theory. And there should be a fluctuation
corresponding to the tachyon field on the non-BPS D$p$-brane.
Note that when $p$ is odd, there is a chirality operator
$\gamma_*$ which anti-commutes with $\gamma_\alpha$
($\alpha=0,1,\dots,p$) and normalized as $\gamma_*^2=1$,
\begin{eqnarray}
\{\gamma^\alpha,\gamma_*\}&=&0,~~~(\alpha=0,1,\dots,p),\\
\gamma_*^2&=&1.
\end{eqnarray}
We can add a fluctuation proportional to $\gamma_*$ in
(\ref{AinT}) as
\begin{eqnarray}
T&=&u\,\sum_{\alpha=0}^p
(\wh{p}_\alpha-iA_\alpha(\wh{x})) \otimes \gamma^\alpha
+t(\wh x)\otimes\gamma_*.
\label{Tfluc0}
\end{eqnarray}
which implies
\begin{eqnarray}
T^2=u^2(\wh{p}_\alpha-iA_\alpha)^2-iu
D_\alpha t\otimes\gamma^\alpha\gamma_*+ t^2
-\frac{u^2}{2}F_{\alpha\beta}\gamma^{\alpha\beta}.
\end{eqnarray}
Then, (\ref{M2}) is modified as
\begin{eqnarray}
M
&=&-(i\wh x^\alpha P_\alpha+i\phi^i P_i+\half[\phi^i,\phi^j]
\Pi_i\Pi_j
+u^2(\wh p_\alpha-iA_\alpha)^2+t^2) I\nn\\
&&
-{[}\phi^i,t{]}\Pi_i\Gamma_*
+iuD_\alpha t\,\Gamma^\alpha\Gamma_*
-iu(\Pi_\alpha+ D_\alpha\phi^i \Pi_i)\Gamma^\alpha
+\frac{u^2}{2}F_{\alpha\beta}\Gamma^{\alpha\beta},
\end{eqnarray}
where $\Gamma_*={~~~\gamma_*\choose \gamma_*~~~}$.

It is straightforward to repeat the argument in section
\ref{gaugefield} and we obtain
\begin{eqnarray}
e^{-S_b(\Phi^\mu,T)}\ket{B(-1);\pm}
\ra e^{-S_b(A_\alpha,\phi^i,t\,)}\ket{Bp;\pm}
\label{nonDp1}
\end{eqnarray}
in the $u\ra\infty$ limit. Here
the boundary interaction is given by
\begin{eqnarray}
e^{-S_b(A_\alpha,\Phi^i,T)}
=\int d\eta
\trP \exp\left\{
\int d\sigma\left(\frac{1}{4}\dot\eta\eta
+\cA+\cF-T^2+i\cD T\eta
\right)
\right\},
\label{nonDp}
\end{eqnarray}
where $\cA$ and $\cF$ are
those defined in (\ref{cA}) and (\ref{cF}), respectively,
and $\cD T$ is given by
\begin{eqnarray}
\cD T=D_\alpha T\,\Psi^\alpha+i{[}\Phi^i,T{]}\,\Pi_i,
\end{eqnarray}
which is obtained by setting $\wt A_\alpha=A_\alpha$
and $\wt\Phi^i=\Phi^i$ in (\ref{cDT}).
This boundary interaction (\ref{nonDp}) is exactly
what we expect for non-BPS D$p$-branes
as explained in section \ref{bdryrev}.

When $p$ is even, we can similarly construct
D$p$-\AD$p$ system by replacing $\gamma^\alpha$
with $\sigma_3\otimes\gamma^\alpha$ in (\ref{sol2-1}).
Then, $\sigma_1\otimes 1$ and $\sigma_2\otimes 1$
play the same role as $\gamma_*$ above.
In this case,
(\ref{Phifluc0}) and (\ref{Tfluc0}) are replaced as
\begin{eqnarray}
\Phi^i&=&\mat{\phi^i(\wh x),,,\wt\phi^i(\wh x)},\\
T&=&\mat{u
(\wh{p}_\alpha-iA_\alpha(\wh{x})) \otimes \gamma^\alpha,
t(\wh x),t(\wh x)^\dag,
-u(\wh{p}_\alpha-i\wt A_\alpha(\wh{x}))
\otimes \gamma^\alpha},
\end{eqnarray}
and the same argument as above precisely implies
the boundary state for D$p$-\AD$p$ system with
boundary interaction given in (\ref{bint})--(\ref{cDT}).

\subsection{Superfield formulation and general fluctuations}
\label{general}

In section \ref{gaugefield}, \ref{scalarfield} and
\ref{tachyonfield}, we have shown that the boundary state
with the boundary interaction representing D$p$-brane
with the fields on it is correctly reproduced from
the non-BPS D-instanton system by considering the fluctuations
$A_\alpha$, $\phi^i$ and $t$ of the form
\begin{eqnarray}
T&=&u\,\sum_{\alpha=0}^p
(\wh{p}_\alpha-iA_\alpha(\wh{x})) \otimes \gamma^\alpha
+t(\wh x)\otimes\gamma_*,
\label{Tfluc}
\\
\Phi^{\alpha} &=& \wh{x}^{\alpha}\otimes 1
\quad
(\alpha=0, \dots, p)\,,
\qquad
\Phi^i=\phi^i(\wh x)\otimes 1
\quad
(i=p+1, \dots, 9).
\label{Phifluc}
\end{eqnarray}
around the D-brane solution
(\ref{sol2-1}, \ref{sol2-2}).
Here $A_\alpha$, $\phi^i$ and $t$
correspond to the gauge, scalar and tachyon fields
on the D$p$-brane, where the tachyon exists for the
cases with odd $p$.
One might think it a little ad hoc,
since we have chosen the fluctuation by hand and shown that they
correspond to the fields on the D$p$-brane, one by one.
In principle, we could consider any fluctuations around the D-brane
solution and there is no reason to restrict them as above
in the first place.
In this subsection, we consider more general fluctuations
and show that the fluctuations other than those considered
above are essentially irrelevant in the limit
$u\ra\infty$.

Before considering fluctuations around the D$p$-brane solution,
we develop an efficient way to analyze it using superfield.
In (\ref{super}), we introduced a superfield description of
the boundary interaction. $\bG^I(\sigma)=\eta^I(\sigma)+\theta\,F^I(\sigma)$
are the fermionic superfields which correspond to gamma matrices $\Gamma^I$.
In fact, the kinetic term for $\bG^I$ implies that
$\eta^I$ satisfy
\begin{eqnarray}
\{\wh\eta^I,\wh\eta^J\}=2\delta^{IJ}
\end{eqnarray}
in the operator formulation,
which is the same anti-commutation relation
as the gamma matrices.
This is why we expanded the matrix $\bM$ by the gamma matrices
in (\ref{gammaexp}).
Now, let us suppose that $\bM^{I_1\dots I_k}$ in (\ref{gammaexp})
are operators acting on a Hilbert
space $\cH$ and can be expressed as (Weyl ordered) polynomials of
the operators $\wh x^\alpha$ and $\wh p_\alpha$, which satisfy the
canonical commutation relation (\ref{ccr}).
Then, following the analogous argument,
we can rewrite the boundary interaction (\ref{super})
in the path integral formulation
introducing superfields
$\bx^\alpha(\sigma)=x^\alpha(\sigma)+i\theta\psi^\alpha(\sigma)$
and
$\bp_\alpha(\sigma)=p_\alpha(\sigma)+i\theta\pi_\alpha(\sigma)$,
which correspond to the operators
$\wh x^\alpha$ and $\wh p_\alpha$, respectively;
\begin{eqnarray}
e^{-S_b}&=&\int[d\bG^I][d\bx^\alpha][d\bp_\alpha]
\,\exp\left\{
\int d\wh\sigma \left(
\frac{1}{4}\bG^I D\bG^I+i\bp_\alpha D\bx^\alpha
\right.\right.\nn\\
&&
~~\left.\left.\frac{}{}
+\sum_{k=0}^{2m} \bM^{I_1\cdots I_k}(\bx,\bp)
\bG^{I_1}\cdots\bG^{I_k}
\right)\right\}.
\label{super3}
\end{eqnarray}
We will give a formal argument to show the equivalence between
(\ref{super3}) and (\ref{supertrP}) in Appendix \ref{Oppath}.

Then, let us again consider the non-BPS D-instantons in type IIA string
theory, in which the boundary interaction is given by (\ref{super3}) with
\begin{eqnarray}
\bM=\mat{-i\Phi^\mu\bP_\mu,T,T,-i\Phi^\mu\bP_\mu}.
\label{superKM}
\end{eqnarray}
To see how this prescription works,
we first reexamine the configuration (\ref{Tfluc}, \ref{Phifluc}).
Inserting (\ref{Tfluc}, \ref{Phifluc}) into (\ref{superKM}),
we obtain
\begin{eqnarray}
\bM(\wh x,\wh p)=-i\wh x^\alpha\bP_\alpha-i\phi^i(\wh x)\bP_i+
 u(\wh p_\alpha-iA_\alpha(\wh x))\Gamma^\alpha+t(\wh x)\Gamma_*.
\end{eqnarray}
{}The boundary interaction (\ref{super3}) is now
\begin{eqnarray}
e^{-S_b(\Phi^\mu,T)}&=&\int[d\bG^I][d\bx^\alpha][d\bp_\alpha]
\trwhP
\exp\left\{
\int d\wh\sigma \left(
\frac{1}{4}\bG^I D\bG^I+i\bp_\alpha D\bx^\alpha
\right.\right.\nn\\
&&\left.\left.
~~-i\bx^\alpha\bP_\alpha-i\phi^i(\bx)\bP_i
+u(\bp_\alpha-iA_\alpha(\bx))\bG^\alpha+t(\bx)\bG_*
\frac{}{}\right)\right\},
\end{eqnarray}
where $\bG^I=(\bG^\alpha,\bG_*)$.
We can perform the path integral over $\bp_\alpha$
which implies
\begin{eqnarray}
iD\bx^\alpha+u\bG^\alpha=0,
\label{Dx}
\end{eqnarray}
and integrating out $\bG^\alpha$, we obtain
\begin{eqnarray}
e^{-S_b(\Phi^\mu,T)}&=&
\int[d\bG_*][d\bx^\alpha]
\trwhP\exp\left\{
\int d\wh\sigma \left(
-\frac{1}{4u^2}D\bx^\alpha D^2 \bx^\alpha
+\frac{1}{4}\bG_* D\bG_*
\right.\right.\nn\\
&&\left.\left.
~~-i\bx^\alpha\bP_\alpha-i\phi^i(\bx)\bP_i
-A_\alpha(\bx)D\bx^\alpha+t(\bx)\bG_*
\frac{}{}\right)\right\}.
\label{superbdry}
\end{eqnarray}
The first term $\frac{1}{4u^2}D\bx^\alpha D^2\bx^\alpha$
in the integrand of (\ref{superbdry}) drops in the limit
$u\ra\infty$ and we have
\begin{eqnarray}
e^{-S_b(\Phi^\mu,T)}\ket{B(-1);\pm}
&\ra&
\int[d\bG_*]\,
\trwhP\exp\left\{
\int d\wh\sigma \,\bM(A_\alpha,\phi^i,t\,)\right\}
\ket{Bp;\pm}\label{nonDp2}
\end{eqnarray}
where
$\bM(A_\alpha,\phi^i,t)$ is
\begin{eqnarray}
\bM(A_\alpha,\phi^i,t)=\frac{1}{4}\bG_* D\bG_*-i\phi^i(\bx)\bP_i
-A_\alpha(\bx)D\bx^\alpha+t(\bx)\bG_*.
\end{eqnarray}
The boundary interaction in the right hand side of
 (\ref{nonDp2}) is nothing but
that for the D$p$-branes (non-BPS or BPS for odd or even $p$, respectively),
which is
given by (\ref{bdrybM}) with the constraints $A_\alpha=\wt A_\alpha$,
$\Phi^i=\wt\Phi^i$ and $T^\dag=T$, as expected.
Therefore, we have reproduced the same result as (\ref{nonDp1}).
However, the proof using superfields becomes much simpler than
that given in the previous subsections.

Next, let us consider general fluctuations which do not
involve the operator $\wh p_\alpha$.
Possible fluctuations can be written as
\begin{eqnarray}
T&=&u\,\wh{p}_\alpha\otimes \gamma^\alpha
+\sum_{k:odd} t_{\alpha_1,\dots,\alpha_k}(\wh x)
\otimes\gamma^{\alpha_1,\dots,\alpha_k}
+\sum_{k:even} t_{*\,\alpha_1,\dots,\alpha_k}(\wh x)
\otimes\gamma_*\gamma^{\alpha_1,\dots,\alpha_k},\nn\\
\\
\Phi^{\alpha} &=& \wh{x}^{\alpha} \otimes {\rm 1}+
\sum_{k:even}\phi^\alpha_{\alpha_1,\dots,\alpha_k}(\wh x)
\otimes \gamma^{\alpha_1,\dots,\alpha_k}
+\sum_{k:odd}\phi^\alpha_{*\,\alpha_1,\dots,\alpha_k}(\wh x)
\otimes \gamma_*\gamma^{\alpha_1,\dots,\alpha_k},
\\
\Phi^i&=&
\sum_{k:even}\phi^i_{\alpha_1,\dots,\alpha_k}(\wh x)
\otimes \gamma^{\alpha_1,\dots,\alpha_k}
+\sum_{k:odd}\phi^i_{*\,\alpha_1,\dots,\alpha_k}(\wh x)
\otimes \gamma_*\gamma^{\alpha_1,\dots,\alpha_k}.
\end{eqnarray}
Here $\gamma_*\propto\gamma^0\gamma^1\cdots\gamma^p$
is a product of even number of $\gamma^\alpha$
for odd $p$. When $p$ is even, since we have
$\gamma^0\gamma^1\cdots\gamma^p\propto 1$,
we can simply drop the terms which involve
$\gamma_*$ in these equations.

Then the matrix (\ref{superKM}) becomes
\begin{eqnarray}
\bM(\wh x,\wh p)
&=&-i\wh x^\alpha\bP_\alpha+u\wh p_\alpha\Gamma^\alpha\nn\\
&&+\sum_{k:odd} t_{\alpha_1,\dots,\alpha_k}(\wh x)
\Gamma^{\alpha_1,\dots,\alpha_k}
+\sum_{k:even} t_{*\,\alpha_1,\dots,\alpha_k}(\wh x)
\Gamma_*\Gamma^{\alpha_1,\dots,\alpha_k}\nn\\
&&-i\sum_{k:even} \phi_{\alpha_1,\dots,\alpha_k}^\mu(\wh x)
\Gamma^{\alpha_1,\dots,\alpha_k}\bP_\mu
-i\sum_{k:odd} \phi_{*\,\alpha_1,\dots,\alpha_k}^\mu(\wh x)
\Gamma_*\Gamma^{\alpha_1,\dots,\alpha_k}\bP_\mu.\nn\\
\label{genflucM}
\end{eqnarray}
Performing the path integral over $\bp_\alpha$, we
again obtain (\ref{Dx}) and,
as a result, the vertex operators
\begin{eqnarray}
&&
\sum_{k:odd}\frac{1}{u^k} t_{\alpha_1,\dots,\alpha_k}
 D\bX^{\alpha_1}\cdots D\bX^{\alpha_k},~
\sum_{k:even}\frac{1}{u^k} t_{*\,\alpha_1,\dots,\alpha_k}
 D\bX^{\alpha_1}\cdots D\bX^{\alpha_k},\nn\\
&&
\sum_{k:even}\frac{1}{u^k}\phi^\mu_{\alpha_1,\dots,\alpha_k}
 D\bX^{\alpha_1}\cdots D\bX^{\alpha_k}\bP_\mu,~
\sum_{k:odd}\frac{1}{u^k}\phi^\mu_{*\,\alpha_1,\dots,\alpha_k}
 D\bX^{\alpha_1}\cdots D\bX^{\alpha_k}\bP_\mu,\nn\\
\end{eqnarray}
are turned on in the boundary interaction.
In order for these fields to survive in the $u\ra\infty$ limit,
$t_{\alpha_1\dots\alpha_k}$, $t_{*\,\alpha_1\dots\alpha_k}$,
$\phi^\mu_{\alpha_1\dots\alpha_k}$ and
$\phi^\mu_{*\,\alpha_1\dots\alpha_k}$,
should be of order $u^k$.
If we allow only such fluctuations that do not increase
more than the D$p$-brane solution itself in the $u\ra\infty$
limit,
that is,
\begin{eqnarray}
T&=&u\,\sum_{\alpha=0}^p
\wh{p}_\alpha \otimes \gamma^\alpha+\cO(u^1),
\\
\Phi^{\alpha} &=& \wh{x}^{\alpha} \otimes {\rm 1}
+\cO(u^0),~~~~~
(\alpha=0, \dots, p), \\
\Phi^i&=&\cO(u^0),~~~~~
(i=p+1, \dots, 9),
\end{eqnarray}
the only relevant fluctuations are
\begin{eqnarray}
T&=&u\,\sum_{\alpha=0}^p
(\wh{p}_\alpha-iA_\alpha(\wh x)) \otimes \gamma^\alpha
+t_*(\wh x)\otimes\gamma_*,
\label{relevant1}
\\
\Phi^{\alpha} &=& \wh{x}^{\alpha} \otimes 1
+\phi^\alpha(\wh x)\otimes 1,~~~~~(\alpha=0, \dots, p),
\label{repara} \\
\Phi^i&=&\phi^i(\wh x)\otimes 1,~~~~~(i=p+1, \dots, 9),
\label{relevant2}
\end{eqnarray}
where we have set $t_\alpha=-iuA_\alpha$.

The fluctuation $\phi^\alpha$ in (\ref{repara})
represent the non-Abelian generalization of the
reparametrization of the world-volume coordinates.
When the fluctuations are Abelian, $\phi^\alpha$ can be absorbed
by the gauge transformation \cite{AsSuTe}. In fact,
the gauge transformation (\ref{gaugetr})
with $U=\exp(i\{\wh p_\alpha, \lambda^\alpha(\wh x)\}/2)$ implies
\begin{eqnarray}
 U \wh x^\alpha U^{-1}= \wh x^\alpha+\lambda^\alpha(\wh x)
 +{\cal O}(\lambda^2),
\label{reparatr}
\end{eqnarray}
and we can always set $\phi^\alpha=0$ by choosing $\lambda^\alpha$
appropriately.
This comes from the reparametrization invariance
of the D$p$-brane world-volume, which is included
in the gauge symmetry in the K-matrix theory \cite{AsSuTe}.
For the non-Abelian cases, we cannot always
set $\phi^\alpha=0$. For example, if a commutator
$[\phi^\alpha,\phi^\beta]$ is non-zero, there is no way
to set this quantity to zero by unitary transformation.
We can try the non-Abelian generalization
of the gauge transformation (\ref{reparatr})
to absorb the fluctuations, but
in this case, the ${\cal O}(\lambda^2)$ terms involve $\wh p_\alpha$
and the $\wh p_\alpha$ dependent fluctuations are induced.
Note also that even in the Abelian cases, the transformation
(\ref{reparatr}) will induce
 $\wh p_\alpha$ dependent fluctuations in the tachyon
operator (\ref{relevant1}), which corresponds to the
transformation of vielbein.

So far we have restricted the fluctuations to be
independent of the operator $\wh p_\alpha$.
It is also important to see what happens if we include
the $\wh p_\alpha$ dependent fluctuations.
We can apply the above prescription at least
when the fluctuations are linear with respect to $\wh p_\alpha$.
Let us add such fluctuations to (\ref{relevant1})--(\ref{relevant2})
and consider
\begin{eqnarray}
T&=&u\,e^\alpha_{~\beta}(\wh x) \,\wh D_\alpha
\otimes \gamma^\beta
+(t_*(\wh x)+t_*^\alpha(\wh x)\wh D_\alpha)\otimes\gamma_*
\\
\Phi^{\alpha} &=&
(\wh x^\alpha+\phi^\alpha(\wh x)+\phi^{\alpha\beta}(\wh x)\wh D_\beta)
\otimes 1,~~~~~~(\alpha=0, \dots, p),
\\
\Phi^i&=&(\phi^i(\wh x)+\phi^{i\alpha}(\wh x)\wh D_\alpha)\otimes 1,
~~~~~(i=p+1, \dots, 9),
\end{eqnarray}
where $\wh D_\alpha=\wh{p}_\alpha-iA_\alpha(\wh x)$ and
the products of $\wh x^\alpha$ and $\wh p_\alpha$ are understood
as Weyl ordered products. We have ignored higher terms in the gamma
matrix expansion since they do not contribute in the limit
$u\ra\infty$ from the argument given above.
Here we consider the Abelian case for simplicity.
The boundary interaction is again given by
inserting this configuration into (\ref{super3}) and (\ref{superKM});
\begin{eqnarray}
e^{-S_b}&=&\int[d\bG^I][d\bx^\alpha][d\bp_\alpha]\,\exp\left\{
\int d\wh\sigma\left(\frac{1}{4}\bG^I D\bG^I+i\bp_\alpha D\bx^\alpha
+\bM(\bx,\bp)\right)\right\}
\label{bdryxp}
\end{eqnarray}
\begin{eqnarray}
\bM(\bx,\bp)&=&-i
\left(\bx^\alpha+\phi^\alpha(\bx)+
\phi^{\alpha\beta}(\bx)(\bp_\beta-iA_\beta(\bx))\right)\bP_\alpha\nn\\
&&-i\left(\phi^i(\bx)+\phi^{i\alpha}(\bx)(\bp_\alpha-iA_\alpha(\bx))
\right)\bP_i\nn\\
&&+u\, e^\alpha_{~\beta}(\bx)\left(\bp_\alpha-iA_\alpha(\bx)\right)\bG^\beta+
\left(t_*(\bx)+t_*^\alpha(\bx)(\bp_\alpha-iA_\alpha(\bx)
\right)\bG_*.
\end{eqnarray}
Performing the path integral with respect to $\bp_\alpha$,
we obtain a delta-functional which imposes the condition
\footnote{
To be more precise,
we should understand the operators $\bP_\beta$ and
$\bP_i$ as their eigen functions by using (\ref{Pcoh}).
}
\begin{eqnarray}
iD\bx^\alpha+u\,e^\alpha_{~\beta}(\bx)\bG^\beta+
t_*^\alpha(\bx)\bG_*-i\phi^{\beta\alpha}(\bx)\bP_\beta
-i\phi^{i\alpha}(\bx)\bP_i=0.
\end{eqnarray}
Then, integrating out $\bG^\beta$,
 we can easily see that all the new degrees of freedom
($i.e.$ $e^\alpha_{~\beta}$, $t_*^\alpha$, $\phi^{\alpha\beta}$ and
$\phi^{i\alpha}$)
will be dropped in the limit $u\ra\infty$.
Therefore, there is no effect from the fluctuations which depend
linearly on $\wh p_\alpha$.
We do not proceed to a systematic
analysis including higher order terms
with respect to $\wh p_\alpha$ for a technical reason.
But, we expect that there are essentially no new degrees of
freedom. Roughly speaking, even if there are such higher
order terms, the largest contribution in the boundary interaction
(\ref{bdryxp}) will be
$u\, e^\alpha_{\,\beta}(\bx)(\bp_\alpha-iA_\alpha(\bx))\bG^\beta$
in the $u\ra\infty$ limit. This will imply the
condition $\bp_\alpha-iA_\alpha(\bx)=0$, from which we can
eliminate all the $\wh p_\alpha$ dependent fluctuations.

\subsection{CS-term and the index theorem}
\label{cs-term}

Here we consider the Chern-Simons term.
Since we have already shown that the D$p$-brane
boundary state can be obtained by considering
fluctuations around the D$p$-brane solution
in K-matrix theory, it is now clear that
the CS-term of D$p$-brane can be
obtained from that of K-matrix theory.\footnote{
We can also obtain, for example, DBI action from K-matrix theory
as done in \cite{Te} explicitly for transverse scalar
fields of the D$p$-brane
although we do not discuss it in this paper.}
More explicitly, (\ref{CS}) and (\ref{nonDp1}) imply
\begin{eqnarray}
S^{D(-1)}_{CS}(C,\Phi^\mu,T)=
S^{Dp}_{CS}(C,A_\mu,\phi^i,t),
\label{CSequiv}
\end{eqnarray}
inserting the configuration (\ref{Tfluc})
and (\ref{Phifluc}) in the limit $u\ra\infty$.
As we can see in (\ref{CSequiv}), the left hand side
is written by operators $\Phi^\mu$ and $T$ acting on
a Hilbert space, while the right hand side is written
in terms of the fields $A_\mu$, $\phi^i$ and $t$
on the D$p$-brane.
For example, if we consider BPS D$p$-branes,
in which $t$ doesn't exist,
and turn off the scalar fields $\phi^i$,
the CS-term for the D$p$-brane takes the usual form
\begin{eqnarray}
S^{Dp}_{CS}(C,A_\mu)=\mu_p\int_{Dp}C\wedge \tr e^{2\pi F},
\end{eqnarray}
up to derivative corrections which do not contribute
to the D-brane charges.
Comparing the coupling to the RR-fields $C$ in the
both sides of (\ref{CSequiv}), we
obtain some interesting formulae which relate
some quantity written in the operator theory
with those written in the gauge theory.
\footnote{
Similar strategy was used in the context of
non-commutative gauge theory in \cite{OkOo}-\cite{LiMi},
in which Seiberg-Witten map is derived by
comparing the RR-coupling in commutative and
non-commutative description of a gauge theory.
}
In particular, as we will see below,
this observation naturally
leads us to obtain a physical
derivation of Atiyah-Singer index theorem.

In the following, we will first derive general CS-terms
in our formulation. It would be instructive to
calculate the CS-term and show the equivalence
(\ref{CSequiv}) explicitly. Then, we will give
the physical interpretation of Atiyah-Singer index theorem
as well as its generalization.

\subsubsection{General CS-terms}
\label{genCS}

As we briefly mentioned in (\ref{CS}), the CS-term for the D$p$-brane
can be written as
\begin{eqnarray}
S_{CS}^{Dp}=\bra{C}\Str {\rm P}\,e^{\int d\sigma M}
\ket{Bp;+}_{\rm RR},
\label{CSDp}
\end{eqnarray}
using the boundary interaction of the form (\ref{supertrace}).
Here we concentrate on the lowest order term in
the derivative expansion. Then, it is known that only zero modes
contribute in the calculation because of the supersymmetry
(See, for example,  \cite{DiVeFrLeLi,Wy}).
Let $\psi_1^\mu$ and $i\psi_2^\mu/2\pi$ be the zero modes
of $\Psi^\mu$ and $\Pi^\mu$, respectively, satisfying
\begin{eqnarray}
\{\psi_1^\mu,\psi_1^\nu\}=\{\psi_2^\mu,\psi_2^\nu\}=0,~~~
\{\psi_1^\mu,\psi_2^\nu\}=\delta^{\mu\nu}.
\label{car}
\end{eqnarray}

Then, inserting the zero mode part of the expressions
(\ref{cA})--(\ref{cDT}) and the boundary state (\ref{bpm})
into (\ref{CSDp}),
we obtain
\begin{eqnarray}
S^{Dp}_{CS}=\mu_p
\int\prod_{\beta=0}^p dx^\beta d\psi_1^\beta
\bra{\psi_2^i=0}
 \int d^{9-p}k\,
 C(x^\alpha,k_i,\psi_1^\mu)\,
 J(x^\alpha,k_i,\psi_1^\alpha,\psi_2^i)
\ket{\psi_1^i=0},\nn\\
\label{DpCS}
\end{eqnarray}
where
\begin{eqnarray}
C(x^\mu,\psi_1^\mu)&=&\sum_n
C_{\mu_1\cdots\mu_n}(x^\mu)\,\psi_1^{\mu_1}\cdots\psi_1^{\mu_n}\\
&=&\int d^{9-p}k\,e^{ix^jk_j} C(x^\alpha,k_i,\psi_1^\mu)
\end{eqnarray}
and
\begin{eqnarray}
J(x^\alpha,k_i,\psi_1^\alpha,\psi_2^i)
=\Str\left( e^{-ik_i\wh\Phi^i+2\pi \wh\cF}\right),
\label{J}
\end{eqnarray}
\begin{eqnarray}
\wh\Phi^i=\mat{\Phi^i,,,\wt\Phi^i},~~~
\wh\cF=\mat{-TT^\dag+\cF,\cD T,
\cD T^\dag,-T^\dag T+\wt\cF},
\end{eqnarray}
\begin{eqnarray}
\cF&=&\frac{1}{2}F_{\alpha\beta}\psi_1^\alpha\psi_1^\beta-
\frac{1}{2\pi}(\del_\alpha\Phi^i+[A_\alpha,\Phi^i])\psi_1^\alpha\psi_2^i
+\frac{1}{8\pi^2}[\Phi^i,\Phi^j]\psi_2^i\psi_2^j,\\
\wt\cF&=&\frac{1}{2}\wt F_{\alpha\beta}\psi_1^\alpha\psi_1^\beta-
\frac{1}{2\pi}
(\del_\alpha\wt\Phi^i+[\wt A_\alpha,\wt\Phi^i])\psi_1^\alpha\psi_2^i
+\frac{1}{8\pi^2}[\wt\Phi^i,\wt\Phi^j]\psi_2^i\psi_2^j,\\
\cD T&=&i(\del_\alpha T+A_\alpha T-T\wt A_\alpha)\psi_1^\alpha
-\frac{i}{2\pi}(\Phi^i T-T\wt\Phi^i)\psi_2^i\\
\cD T^\dag&=&i(\del_\alpha T^\dag+\wt A_\alpha T^\dag
-T^\dag A_\alpha)\psi_1^\alpha
-\frac{i}{2\pi}(\wt\Phi^i T^\dag-T^\dag\Phi^i)\psi_2^i.
\end{eqnarray}
Here the indices run as $\mu,\nu=0,\dots,9$,~
 $\alpha,\beta=0,\dots,p$
and $i,j=p+1,\dots,9$.
In (\ref{DpCS}),
the zero modes of the operators
 $X^\alpha$, $P_i$ and $\Psi^\alpha$ in the boundary
 interaction are replaced with their eigen values
 denoted as $x^\alpha$,
$k_i$ and $\psi_1^\alpha$, respectively,
while we still treat $\psi_1^i$ and $\psi_2^i$
as fermionic operators satisfying (\ref{car}).
$\bra{\psi_2^i=0}$ and $\ket{\psi_1^i=0}$
are bra and ket states which are
annihilated by $\psi_2^i$ and $\psi_1^i$, respectively,
satisfying $\bracket{\psi_2^i=0}{\psi_1^i=0}=1$.
$\bra{\psi_2^i=0}$ is originally a zero momentum
Fock vacuum on which RR-states are constructed,
\footnote{
See \cite{DiVecchia et al,DiVeFrLeLi} for more precise argument.
}
while $\ket{\psi_1^i=0}$ is the zero mode
part of the coherent state
$\ket{x^\alpha,x^i=0}\ket{\psi^\alpha,\psi^i=0}_{\rm RR}$.
The normalization factor $\mu_p=1/(2\pi)^{p+1}$
in (\ref{DpCS}) comes from the path integration
with respect to the non-zero modes.
(See Appendix \ref{Normalization}.)

When the scalar fields $\Phi^i$, $\wt\Phi^i$ are
turned off, this CS-term will be reduced to
the well-known form \cite{KeWi,KrLa,TaTeUe}
\begin{eqnarray}
S_{CS}^{Dp}=\mu_p\int C\wedge
\Str e^{2\pi \wh\cF}
\end{eqnarray}
with the supercurvature \cite{Qu}
\begin{eqnarray}
\wh\cF=\mat{-TT^\dag +\frac{1}{2} F_{\alpha\beta}dx^\alpha dx^\beta,
iD_\alpha T\, dx^\alpha,iD_\alpha T^\dag\, dx^\alpha,
-T^\dag T+\frac{1}{2}\wt F_{\alpha\beta}dx^\alpha dx^\beta}.
\end{eqnarray}

For the BPS D$p$-branes, we can rewrite (\ref{DpCS})
in more familiar form.
The CS-term for the BPS D$p$-brane can be obtained by
replacing (\ref{J}) with
\begin{eqnarray}
J(x^\alpha,k_i,\psi_1^\alpha,\psi_2^i)
=\Tr\left(\exp\left\{\,-ik_i\Phi^i+
\pi F_{\alpha\beta}\psi_1^\alpha\psi_1^\beta-
D_\alpha\Phi^i\psi_1^\alpha\psi_2^i
+\frac{1}{4\pi}[\Phi^i,\Phi^j]\psi_2^i\psi_2^j\,
\right\}\right).\nn\\
\label{BPSJ}
\end{eqnarray}
We use the notation
\begin{eqnarray}
C(y^\mu,\psi_1^\mu)\equiv
\int d^{9-p}k\, C(x^\alpha,k_i,\psi_1^\mu)\,e^{-ik_i\Phi^i},
\label{nonC}
\end{eqnarray}
where $y^\mu=(x^\alpha,\Phi^i(x^\alpha))$,
in the following.
This is just the usual Fourier transformation in the case with
a single D$p$-brane.

As a warm up exercise, let us
first explain the case with a single D$p$-brane.
In this case, every field commutes with each other and we obtain
\begin{eqnarray}
S_{CS}^{Dp}&=&\mu_p
\int \prod_{\beta}dx^\beta d\psi_1^\beta
\bra{\psi_2^i=0}C(y^\mu,\psi_1^\mu)
e^{\del_\alpha y^i\psi_1^\alpha\psi^i_2}
\ket{\psi_1^i=0}
e^{\pi F_{\alpha\beta}\psi_1^\alpha\psi_1^\beta}\\
&=&\mu_p
\int \prod_{\beta}dx^\beta d\psi_1^\beta\,
C(y^\mu,\del_\alpha y^\mu\psi_1^\alpha)\,
e^{\pi F_{\alpha\beta}\psi_1^\alpha\psi_1^\beta}.
\end{eqnarray}
Furthermore, note that
$\psi_1^\beta$ anti-commutes with each other
and the integral with respect
to $\psi_1^\beta$ picks up the coefficient
of $\psi_1^0\psi_1^1\cdots \psi_1^p$.
This enables us to regard $\psi_1^\alpha$ as the
1-form $dx^\alpha$ on the world-volume. Thus we reproduce
the CS-term for the BPS D$p$-brane,
\begin{eqnarray}
S_{CS}^{Dp}=\mu_p\int_{Dp}
C(y^\mu,d y^\mu)\wedge
e^{2\pi F},
\end{eqnarray}
where $C(y^\mu,dy^\mu)$ is a pull back of the RR-field
to the D$p$-brane world-volume
and $F=\half F_{\alpha\beta}dx^\alpha \wedge dx^\beta$.

Next, we consider the non-Abelian case.
Note that
when we expand the exponential in (\ref{J}),
the matrices $\Phi^i$, $D_\alpha\Phi^i$,
$[\Phi^i,\Phi^j]$ and $F_{\alpha\beta}$ always
appear in symmetric order.
Therefore, we can forget about the ordering of
these by simply replacing the trace $\Tr$
with the symmetrized trace.
So we can proceed as in the Abelian case
except the contribution of the last term in (\ref{BPSJ}).
By noticing that $\psi_2^i$ can be identified as
$\frac{\del}{\del\psi_1^i}$, we see that
this term simply modifies the RR-field $C$ to
\begin{eqnarray}
C_{[y^\mu,y^\nu]}(y^{\mu},\psi_1^\mu)
\equiv e^{\frac{1}{4\pi}[y^\mu,y^\nu]
\frac{\del}{\del\psi_1^\mu} \wedge \frac{\del}{\del\psi_1^\nu} }
C (y^{\mu},\psi_1^\mu).
\end{eqnarray}
This corresponds to inclusion of Myers term \cite{My}.
Then, the CS-term (\ref{DpCS}) will become
\beq
S^{Dp}_{CS}=\mu_p\int_{Dp} {\rm SymTr} \,\,\left(
C_{[y^{\mu},y^{\nu}]}(y^{\mu},D y^\mu )  \,
\wedge e^{2\pi F}
\right),
\label{BPScs}
\eeq
where $Dy^\mu=D_\alpha y^\mu(x)dx^\alpha$
and ${\rm SymTr}$ denotes the symmetrized trace
with respect to $y^\mu$, $D_\alpha y^\mu$, $F_{\alpha\beta}$ and
$[y^\mu,y^\nu]$.

\subsubsection{Ascent relations of CS-terms}
\label{CSascent}

Now we explicitly show the equivalence (\ref{CSequiv}) between
the CS-term for the D$p$-branes and
that for the unstable D-instanton system
in the background of the D$p$-brane solution,
using the expressions given in the last subsection.
Let us explain this in type IIA string theory.
Our starting point is the CS-term for the non-BPS D-instantons
\cite{AsSuTe,TaTeUe}
which is obtained by considering $p=-1$ case
in (\ref{DpCS});
\beq
S_{CS}^{D(-1)}= \VEV{
\int d^{10}k\,C(k_\mu,\psi_1^\mu) J(k_\mu,\psi_2^\mu)},
\label{CSDin}
\eeq
where $\VEV{\cdots}$ denotes
 $\bra{\psi_2^\mu=0}\cdots\ket{\psi_1^\mu=0}$ and
\beq
J(k,\psi_2)&=& \Str \left(e^{-ik_\mu \Phi^\mu+2\pi\wh\cF} \right),
\label{JZ}\\
\wh\cF &=&
-T^2+\frac{1}{8\pi^2}[\Phi^\mu,\Phi^\nu]\psi_2^\mu\psi_2^\nu
-\frac{i}{2\pi}[\Phi^\mu,T]\psi_2^\mu\sigma_1.
\label{hatF}
\eeq
Here the Pauli matrix $\sigma_1$ is considered as fermionic
matrix which anti-commutes with $\psi_1^\mu$ and $\psi_2^\mu$,
as explained in section \ref{bdryrev}.

Let us consider the solution representing
 BPS D$p$-branes ($p=\mbox{even}$) and the fluctuations
around it. Inserting
(\ref{relevant1})--(\ref{relevant2}) into (\ref{hatF})
we obtain
\begin{eqnarray}
\wh\cF=-u^2(\wh p_\alpha-iA_\alpha(\wh x))^2
+\frac{1}{8\pi^2}[y^\mu(\wh x),y^\nu(\wh x)]\psi_2^\mu\psi_2^\nu
+\frac{u}{2\pi}D_\alpha y^\mu(\wh x)\psi_2^\mu\Gamma^\alpha
+\frac{u^2}{2} F_{\alpha\beta}(\wh x)\Gamma^{\alpha\beta},\nn\\
\end{eqnarray}
Here we have introduced
\beq
y^\mu(x)= (x^\alpha+\phi^\alpha(x), \phi^i(x) ),
\eeq
which is a non-Abelian generalization
of positions of the BPS D$p$-brane world-volume.

Then we rewrite $J(k,\psi_2)$ in path-integral formulation
with respect to $\hat{x},\hat{p}$ and $\Gamma$
as in the previous subsections.
Actually, we have already done the relevant calculations
in section \ref{gaugefield}.
The result is
\begin{eqnarray}
J(k,\psi_2)= \int
[ d x^{\alpha} ] [ d \psi^{\alpha}]\,
\trP \exp \left(-\int d \sigma\,
L(x^\alpha,\psi^\alpha,k_\mu,\psi_2^\mu)
\right),
\label{csresult}
\end{eqnarray}
where the Lagrangian $L(x,\psi,k,\psi_2)$
is given by
\beqa
L(x,\psi,k,\psi_2)
& = & \frac{1}{4u^2} \left(
\dot{x}_\alpha^2+\psi^\alpha \dot{\psi}^\alpha \right)
+\frac{i}{2\pi}y^\mu (x) k_\mu
-\frac{1}{2\pi}D_{\alpha} y^\mu (x) \psi_2^\mu \psi^{\alpha}\CR
&&-\frac{1}{8\pi^2} [ y^\mu(x), y^\nu (x)] \psi_2^\mu \psi_2^\nu
+A_\alpha(x) \dot{x}^{\alpha}
-\frac{1}{2} F_{\alpha \beta}(x) \psi^{\alpha} \psi^{\beta}.
\label{lag}
\eeqa
The trace $\tr$ in (\ref{csresult}) is
taken with respect to the gauge indices of the
resulting BPS D$p$-branes.

The path-integration of
the non-zero modes of $x(\sigma)$ and $\psi(\sigma)$ in (\ref{csresult})
is trivial by the supersymmetry \cite{KrLa}
and we get
\beqa
S_{CS}
 &=&\mu_p
 \left\langle
 \int  d^{10} k \,
d^{p+1} x \, d^{p+1} \psi \,
C(k^{\mu},\psi_1^{\mu} ) \, \times \right.\CR
&&\left.
\tr\,\exp \left(
-i y^{\mu} k_{\mu}
+ D_{\alpha} y^{\mu}(x) \psi_2^\mu \psi^{\alpha}
+\pi F_{\alpha \beta}(x) \psi^{\alpha} \psi^{\beta}
+\frac{1}{4\pi} [ y^\mu(x), y^\nu(x)] \psi_2^\mu \psi_2^\nu
\right)\right\rangle
\nn\\
&=&\mu_p\left\langle
 \int_{Dp}\int d^{10}k\,
  C(k^{\mu},\psi_1^{\mu})\,
\tr\, e^{ \left(-iy^\mu k_\mu
+D_{\alpha} y^{\mu} \psi_2^\mu dx^{\alpha}
+\pi F_{\alpha \beta} dx^{\alpha}\wedge dx^{\beta}
+\frac{1}{4\pi} [ y^\mu, y^\nu] \psi_2^\mu \psi_2^\nu
\right)}\right\rangle\nn\\
&=&\mu_p\int_{Dp} {\rm Sym}\tr \,\,\left(
C_{[y^{\mu},y^{\nu}]}(y^{\mu},D y^\mu )  \,
\wedge e^{2\pi F}
\right).
\label{csresult2}
\eeqa
This correctly reproduces the CS-term for BPS D$p$-branes
(\ref{BPScs}) when we choose the static gauge $\phi^\alpha=0$.

\subsubsection{An interpretation of the index theorem}
\label{index}

Here we give an interpretation of the index theorem
from the point of view of the K-matrix theory.
We will consider  BPS D$p$-branes constructed from
type IIB K-matrix theory, $i.e.$
an unstable system with infinitely many
D-instanton - anti D-instanton pairs
since it is directly related to the well-known Atiyah-Singer index theorem,
though generalization to other
cases is straightforward.

The Chern-Simons term of the
D-instanton - anti D-instanton system
can be written as (\ref{CSDin}), while (\ref{JZ}) is now
replaced with
\beq
J(k,\psi_2)&=& \Str \left(e^{-ik_\mu\wh\Phi^\mu+2\pi\wh\cF}\right),\\
\wh\cF&=&
\mat{-TT^\dag+\frac{1}{8\pi^2}
{[}\Phi^\mu{,}\Phi^\nu{]}\psi_2^\mu\psi_2^\nu
,-\frac{i}{2\pi}(\Phi^\mu T-T\wt\Phi^\mu)\psi_2^\mu
,-\frac{i}{2\pi}(\wt\Phi^{\mu}T^\dag-T^\dag \Phi^{\mu})\psi_2^\mu,
-T^\dag T+\frac{1}{8\pi^2}
{[}\wt\Phi^\mu{,}\wt\Phi^\nu{]}\psi_2^\mu\psi_2^\nu}.
\label{IIBF}
\eeq
Note that (\ref{IIBF}) can also be written as
\begin{eqnarray}
\wh\cF=-Z^2,~~~
Z=\mat{\frac{i}{2\pi}\Phi^\mu\psi_2^\mu,T,T^\dag,
\frac{i}{2\pi}\wt\Phi^\mu\psi_2^\mu}.
\end{eqnarray}
Let us assume that the RR-field is constant ($i.e.$ $k_\mu=0$)
and all the components except for the zero-form part $C_0$ are
zero. Then the CS-term takes very simple form,
since scalar fields do not contribute to it.
Actually, it becomes the index of the tachyon operator,
\begin{eqnarray}
S_{CS}^{D(-1)}=C_0\Str\left(e^{-2\pi Q^2}\right)
=C_0\Tr\left((-1)^F e^{-2\pi Q^2}\right)=C_0\,{\rm Index}\,Q.
\label{indexQ}
\end{eqnarray}
where $(-1)^F=\sigma^3$ and
\beq
Q \equiv \mat{0,T,T^\dag,0}.
\eeq
Since the coupling of the RR $p$-form field represents
D$(p-1)$-brane charge, we conclude that the index of the tachyon
operator is interpreted as the D-instanton charge
\cite{AsSuTe}.
This can also be seen from the fact that
in the case $T$ is a Fredholm operator,
$\dim {\rm Ker}\, TT^\dag$ and $\dim {\rm Ker}\, T^\dag T$
are finite dimensional and
correspond to the number of D-instantons
and anti D-instantons which are not annihilated by
the tachyon condensation, respectively. It follows that
the total D-instanton number is given by
\begin{eqnarray}
\dim {\rm Ker}\, TT^\dag-\dim {\rm Ker}\, T^\dag T
={\rm Index}\,Q,
\end{eqnarray}
which is another definition of the ${\rm Index}\,Q$.

Let us next consider the D-instanton charge in
the presence of BPS D$p$-branes.
The tachyon configuration representing
D$p$-brane in type IIB K-matrix theory is also
given as (\ref{AinT}) and we have
\beq
Q=u \sum_{\alpha=0}^p
(\wh{p}_\alpha -iA_\alpha(\wh{x}) )\Gamma^{\alpha}
\equiv -iu \Dslash~,
\label{QDirac}
\eeq
where $\Gamma^{\alpha}$ ($\alpha=0,\dots,p$)
are $SO(p+1)$ gamma matrices ($p={\rm odd}$).
Note that $\sigma_3$ can be regarded as
the chirality operator.
Therefore, according to the above observation,
the D-instanton charge in the presence of
the BPS D$p$-branes is just the index of the usual
Dirac operator $\Dslash$ on the world-volume of the
D$p$-branes in the presence of an electro-magnetic background.

On the other hand, as we have explicitly
shown in the previous subsection,
the CS-term for the D-instanton
$S_{CS}^{D(-1)}$ in the background of D$p$-brane
is identical to the CS-term of D$p$-brane $S_{CS}^{Dp}$.
In our case with $C=C_0={\rm constant}$,
the CS-term of D$p$-brane $S_{CS}^{Dp}$ is given as
\beq
S_{CS}^{Dp}=\mu_p\int_{Dp} C \wedge \tr e^{2\pi F}
=C_0 \int_{Dp} \tr e^{F/2\pi},
\label{indexth}
\eeq
where ${\rm tr}$ means the trace for the gauge indices.
Comparing this with (\ref{indexQ}), we obtain
\begin{eqnarray}
{\rm Index}\,(-i\Dslash)=\int_{Dp}\tr e^{F/2\pi}.
\label{AtySin}
\end{eqnarray}
This is nothing but the Atiyah-Singer index
theorem \cite{Atiyah-Singer}.
It is quite interesting that we have naturally reached
this result considering D-brane physics.
The physical interpretation is now clear.
The Dirac operator is the tachyon operator
which represent the D$p$-brane in the
D-instanton - anti D-instanton system
and its index gives the D-instanton charge.
The D-instanton can also be constructed as the instanton
configuration in the gauge theory on the D$p$-brane
world-volume and the instanton number is given by
the Chern number of the gauge bundle.
And the two descriptions actually agree
 as expected.

We can also include a nontrivial internal metric in the Dirac operator
(\ref{QDirac}) as
\beq
Q=-iu\Dslash=-iu \sum_{\alpha=0}^p
e_A^\alpha(\wh x)\left(\wh{\del}_\alpha + \omega_\alpha(\wh x)+
A_\alpha(\wh{x})\right)\,\Gamma^A,
\label{curvedT}
\eeq
where $\wh\del_\alpha=i\wh p_\alpha$
and $\omega_\alpha$ is the spin connection defined by
the vielbein $e_A^\alpha$.
As explained in \cite{AsSuTe}, this vielbein defines
an internal metric $g_{\alpha\beta}=e_\alpha^Ae_\beta^A$,
which is immanent in the spectral triple representing
the D-brane configuration.
Then, repeating the argument given in section \ref{CSascent},
we obtain
\begin{eqnarray}
S_{CS}^{D(-1)}=
C_0\int[dx][d\psi]\TrP\exp\left(-\int d\sigma\,L(x^\alpha,\psi^\alpha)
\right),
\label{curvedCS}
\end{eqnarray}
where
\begin{eqnarray}
L(x^\alpha,\psi^\alpha)&=&
\frac{1}{4u^2}\,
g_{\alpha\beta}(x)
\left(\dot x^\alpha \dot x^\beta+
\psi^\alpha\nabla_\sigma\psi^\beta
\right)\nn\\
&&+A_\alpha(x)\dot x^\alpha
-\half F_{\alpha\beta}(x)\psi^\alpha\psi^\beta,
\label{curvedLag}
\end{eqnarray}
where
$\nabla_\sigma\psi^\alpha=\dot\psi^\alpha+
\dot x^\beta\Gamma^\alpha_{\beta\gamma}\psi^\gamma$.
This system is nothing but the supersymmetric
quantum mechanics which was used to derive the Atiyah-Singer index
theorem in \cite{Al,FrWi}.

In this case, since
we are now interested in the dependence of the curvature
in CS-term,
we have to calculate the derivative corrections,
in which the path integral with respect to non-zero modes
will contribute.
Fortunately, the relevant calculation was
essentially done in \cite{Al,FrWi}
and the result is exactly what we expect from the index theorem;
\beq
S_{CS}^{D(-1)}=C_0 \int_{Dp} \wh{A}(R) \wedge \Tr e^{F/{2\pi}},
\label{indexg}
\eeq
where $\wh{A}(R)$ is the A-roof genus written in terms of the
internal metric introduced in (\ref{curvedT}).

Note that above calculation \cite{Al,FrWi} is consistently done in $u\ra 0$
limit, which is also taken in the heat kernel method.
Since the index (\ref{indexQ}) does not depend
on the parameter $u$, we can take this
rather unphysical limit.
This limit can also be thought of as the zero slope limit
 $\alpha'\ra 0$. If we consider the parameter $u$ as
 a dimensionless parameter
and recover the $\alpha'$ dependence, the D-brane configuration
will become
\beq
T=u\, \sqrt{\frac{\alpha'}{2}}
\left(\wh p_\alpha + A_\alpha(\wh x)\right)\gamma^\alpha,
\,\,\, \Phi^\alpha=\sqrt{\frac{2}{\alpha'}}\,\wh x^\alpha.
\eeq
Here, we treat $T$ and $\Phi^\alpha$ as dimensionless operators.
The normalization is fixed by requiring that the
path integral representation of the operator $\wh x^\alpha$
will have the same
normalization as the closed string coordinate $X^\alpha(\sigma)$.

Then, (\ref{indexQ}) will become
\begin{eqnarray}
S_{CS}^{D(-1)}=C_0\Str\left(
e^{2\pi u^2\alpha'\Dslash^2}
\right),
\end{eqnarray}
and hence the $\alpha'\ra 0$ limit plays the same role as
$u\ra 0$ limit in this case.

However, this is not the end of the story.
The CS-term (\ref{DpCS}) or (\ref{CSDin}) that we have been using is
only reliable for the lowest order in the derivative expansions.
In order to fully incorporate the derivative corrections
of the CS-term, we should have started from (\ref{CSDp}).
The ascent relation for the CS-terms (\ref{CSequiv})
is also valid
including the derivative corrections, since
it has been derived among the boundary states
without any loss of information.
In this case, the fluctuations of the scalar fields
will contribute to the calculation of D-instanton charge density,
and $\alpha'$ cannot be absorbed in the parameter $u$ anymore.
In order to obtain the fluctuation around
D$p$-brane solution, we have to take $u\ra\infty$
keeping $\alpha'$ finite. In the $u\ra\infty$ limit,
the dependence upon the vielbein $e^\alpha_A$ disappears.
However, we still obtain the same formula as (\ref{indexg}),
where the curvature two-form $R$ with respect to
the internal metric is replaced with that defined by
the induced metric on the D$p$-brane world-volume.
The calculation will be summarized in Appendix \ref{AppAroof}.
In this case, we can perform the calculation in
the zero slope limit $\alpha'\ra 0$.
This result is consistent with what we expect in the CS-term
for the D$p$-brane \cite{CS-term}.
\footnote{
In \cite{AlSc,AsSuTe} (see also \cite{MiMo}),
 the CS-term is written by
Todd class $Td(R)$ instead of the A-roof genus
$\wh A(R)$.
Here we have implicitly assumed that there
is a spin structure on the world volume of the D$p$-brane
since we have introduced the gamma matrices on it.
In this case we have $\wh A(R)=Td(R)$ and
there is no discrepancy. We can also treat
the ${\rm Spin}^c$ cases by turning on a $U(1)$ gauge field
associated with the ${\rm Spin}^c$ structure.
}
(See also \cite{Wy} for a related calculation.)
Since the D-instanton charge is quantized,
the coefficient of $C_0$ should not depend on
the continuous parameters $u$ and $\alpha'$,
nor the choice of the metric on the D$p$-brane.
Therefore, the previous result (\ref{indexg})
also gives a correct D-instanton charge
and the interpretation of the index of the tachyon
as the D-instanton charge is still valid,
though it is not the case for the charge density.

We can easily extend the argument given
to the index theorems which involve even or odd superconnections
by considering the solution of D$p$-\AD$p$ pairs
or non-BPS D$p$-branes, respectively.
We could also start from the unstable D$q$-brane system and
construct D$(q+p)$-branes using the ascent relation.
In this case, the tachyon is the $p$-dimensional Dirac operator
parametrized by the world volume coordinates
of the unstable D$q$-brane system.
We can again obtain an index theorem by comparing
the coupling to the RR-fields in the two equivalent
descriptions.
The corresponding index theorem is called
family index theorem, which is related to KK-theory
as expected from the result of \cite{AsSuTe,AsSuTe2}.
It is also possible to generalize the argument to
type I string theory, as we will discuss
in section \ref{typeICSind}.

As we have seen in this subsection,
we can think the index theorem as
a topological aspect of
the ascent relation among the D-brane
boundary states explained in the previous subsections.
In other words, we can think
the D-brane ascent relations as
a kind of generalization of the index theorem.
Actually the essence of the derivation
of the D-brane ascent relation is the equivalence
between path integral and operator formulations
of a supersymmetric quantum mechanics,
which is also used in
the physical proof of the index theorem
given in \cite{FrWi,Al,KrLa}.

\subsection{Application to non-commutative gauge theory}
\label{NCGT}

Before closing this section,
let us make a brief comment on the application to
non-commutative gauge theory.
In this paper, we mainly concentrate on the fluctuations
around the commutative D-brane solutions.
However, our strategy is also applicable to
non-commutative D-branes. In fact, it has been argued in
\cite{Is2,Ok2,Co} that
the equivalence of commutative and non-commutative
descriptions of D-brane in bosonic string theory
can be clearly understood using boundary states.

It would be instructive to show this is also the case
for the superstring case using our superfield formulation.
To be specific, let us consider a non-commutative D$2$-brane
in type IIA string theory.
The non-commutative D$2$-brane can be constructed
from infinitely many D$0$-branes
by turning on non-commutativity between $\Phi^1$ and $\Phi^2$ as
\begin{eqnarray}
\Phi^\alpha=\wh z^\alpha,~~~(\alpha=1,2)
\label{nccoord}
\end{eqnarray}
with $[\wh z^\alpha,\wh z^\beta]=i\Theta^{\alpha\beta}$,
where $\Theta^{\alpha\beta}=-\Theta^{\beta\alpha}$ is
the real constant parameter which represent
the non-commutativity.

Then the boundary interaction is given by
\begin{eqnarray}
e^{-S_b}&=&\int[d\bz^\alpha]\,\exp\left\{
-i\int d\wh\sigma\left(
\Theta^{-1}_{\alpha\beta}\,\bz^\alpha D\bz^\beta+\bz^\alpha\bP_\alpha
\right)\right\},
\end{eqnarray}
where we have introduced the superfields $\bz^\alpha$
corresponding to the operators $\wh z^\alpha$,
and the non-commutativity is represented in their kinetic term.
Therefore, the boundary state will become
\begin{eqnarray}
e^{-S_b}\ket{B0;\pm}=e^{-i\int d\hat\sigma\, \Theta^{-1}_{\alpha\beta}
\bX^\alpha D\bX^\beta}
\ket{B2;\pm},
\end{eqnarray}
which is equivalent to a D$2$-brane boundary state
with constant magnetic flux
$F_{\alpha\beta}=i\Theta^{-1}_{\alpha\beta}$,
which can also be considered as the effect of
background $B$-field
$B_{\alpha\beta}=\Theta^{-1}_{\alpha\beta}$.\cite{Is,Ok}

It is also easy to include the fluctuations around
the solution (\ref{nccoord})
repeating the argument given in section \ref{gaugefield} for
the commutative case.
The local gauge transformation of this system is given by the
unitary transformation  $U(\wh z)$
 which depend on the non-commutative coordinates
$\wh z^\alpha$. Since $\wh z^\alpha$ don't commute with $U(\wh z)$,
 we introduce the non-commutative gauge field
$\wh A_\alpha(\wh z)$ to maintain the local gauge invariance
and replace (\ref{nccoord}) as
\begin{eqnarray}
\Phi^\alpha=\wh z^\alpha+i\Theta^{\alpha\beta}\wh A_\beta(\wh z).
\end{eqnarray}
Then, the boundary state will become
\begin{eqnarray}
\int[d\bz^\alpha]\,\exp\left\{-i
\int d\wh\sigma\left(
\Theta^{-1}_{\alpha\beta}\,\bz^\alpha D\bz^\beta+
\left(\bz^\alpha+i\Theta^{\alpha\beta}\wh A_\beta(\bz)\right)
\bP_\alpha
\right)\right\}
\ket{B0;\pm}.
\label{NC}
\end{eqnarray}
On the other hand, the boundary state of a D2-brane with the gauge field
in the constant $B$-field background is given as
\begin{eqnarray}
\int[d\bx^\alpha]\,\exp\left\{-i
\int d\wh\sigma\left(
B_{\alpha\beta}\,\bx^\alpha D\bx^\beta-i
A_\alpha(\bx)D\bx^\alpha+\bx^\alpha\bP_\alpha
\right)\right\}
\ket{B0;\pm}.
\label{D2inB}
\end{eqnarray}
These two boundary states (\ref{NC}) and (\ref{D2inB})
are equal when $B_{\alpha\beta}=\Theta_{\alpha\beta}^{-1}$
and the gauge fields $\wh A_\alpha$ and $A_\alpha$
related by the Seiberg-Witten map \cite{SW}.
Actually, (\ref{NC}) and (\ref{D2inB})
can be obtained by
simply replacing all the fields with superfields
appearing in the corresponding boundary states
$\ket{\phi}_{\rm NC}$ and $\ket{\cA}$
defined in \cite{Ok2}, respectively,
and it is straightforward to generalize
the argument in \cite{Ok2}
to find the Seiberg-Witten map requiring the equality
of the two boundary states (\ref{NC}) and (\ref{D2inB}).

\section{Generalization to higher dimensional systems}

So far, we have considered the construction of D-branes
in K-matrix theory, which is
the lowest dimensional unstable D-brane system.
The method developed in the previous section can also be
applied to the construction of D-branes in
higher dimensional unstable D-brane systems.
In this section, we generalize the above consideration
to the higher dimensional unstable D-brane systems
and show the general D-brane descent/ascent relations.
First, we consider space-time filling unstable D-brane system
and provide a simple derivation of D-brane descent relations.
Then, we further generalize the argument
to the construction of D$p$-branes from unstable D$q$-brane systems.

\subsection{D-brane descent relations}

Let us consider non-BPS D9-branes in type IIA string theory.
In this case, the configuration representing
D$p$-branes with the fluctuations corresponding to
the massless fields is expected to be
\beq
T=u \gamma^i (x^i-\phi^i(x^\alpha)), \;\;\;
A_\alpha= A_\alpha(x^\alpha), \;\; A_i=0,
\label{d9}
\eeq
where $i=p+1, \ldots, 9$ and $\alpha=0, \ldots, p$
\cite{KuMaMo2,Ho}.
Here we restrict $p=\mbox{even}$ and consider BPS D$p$-branes
for simplicity.
(See the next subsection for the generalization
to more general situations.)
The fluctuation corresponding to the scalar
fields $\phi^i(x^\alpha)$ appear in this way
in the tachyon field, since they correspond to the
Nambu-Goldstone modes for the spontaneous breaking of
the translational symmetry $x^i\ra x^i+\epsilon^i$.

Then, $\bM$ in (\ref{superM}) becomes
\beq
\bM=
u (\bX^i -
\phi^i (\bX^\alpha) )\Gamma^i
-{A_\alpha} (\bX^\alpha) D \bX^\alpha.
\eeq
Using (\ref{BSb2}) and the
boundary interaction of the form (\ref{super}),
we obtain
the non-BPS D9 brane boundary state in the background
(\ref{d9});
\beq
\ket{B9;\pm}_{S_b}
&=& \int [d \bx^\mu]  [d \bG^i]\,\Tr{\wh{\rm P}}\,
\exp \left\{ \int d\wh\sigma \,
\left( \frac{1}{4}\bG^i D\bG^i +
u (\bx^i-\phi^i (\bx^\alpha) )\bG^i\right.\right.\nn\\
&&~~~~~\left.\left.
-A_\alpha(\bx^\alpha)D\bx^\alpha
-i\bx^\mu \bP_\mu
\frac{}{}\right) \right\}
\ket{\bx^\mu=0;\pm},
\label{B9Dp}
\eeq
where $\mu=0, \ldots,9$ and
$\ket{\bx^\mu=0;\pm}=\ket{x^\mu=0}\ket{\psi^\mu=0;\pm}$.
Note that the boundary action is linear with respect to
$\bx^i$ and the path integration of it
gives $\delta(u \bG^i -i\bP_i)$.
This expression is rather formal, since
$\bP_i$ is an operator constructed by
the oscillators $\alpha_m^i$, $\wt\alpha_m^i$, $\Psi_r^\mu$ and
$\wt\Psi_r^\mu$, while $\bG^i$ is an ordinary function.
To be more precise, if we introduce a coherent state
for the momentum operators which satisfy
\begin{eqnarray}
\bP_\mu(\sigma)\ket{\wt\bp_\mu;\pm}=\wt\bp_\mu(\sigma)\ket{\wt\bp_\mu;\pm},
\label{Pcoh}
\end{eqnarray}
\begin{eqnarray}
\ket{\bx^\mu;\pm}=\int[d\wt\bp_\mu]
\,e^{-i\int d\hat\sigma \tilde\bp_\mu\bx^\mu}\ket{\wt\bp_\mu;\pm},
\end{eqnarray}
the formal delta-functional and the arguments below
will be justified.

Then, integrating out $\bG^i$, we obtain
\begin{eqnarray}
\ket{B9;\pm}_{S_b}
&=&\int [d \bx^\alpha]\Tr{\wh{\rm P}}\,
\exp\left\{-\int d\wh\sigma  \left(
\frac{1}{4 u^2} \bP_i D \bP_i+i\phi^i (\bx^\alpha) \bP_i
\right.\right.\nn\\
&&~~~~
\left.\left.\frac{}{}
+A_\alpha (\bx^\alpha) D \bx^\alpha + i\bx^\alpha\bP_\alpha
\right)\right\}
\ket{\bx^\mu=0;\pm}.
\end{eqnarray}
Taking $u \ra \infty $ limit,
we correctly obtain the the boundary state for
the D$p$-brane with the boundary interaction including
all the massless fields on it. This gives a simple proof of the
D-brane descent relations.

As we have demonstrated in section \ref{cs-term},
it is straightforward to apply this result to derive
the CS-term for D$p$-branes from that for an unstable
D9-brane system.
One of the interesting points in our derivation is that
the the gamma matrices $\gamma^i$ are replaced
with the fermionic fields $\eta^i(\sigma)$
(the first component of the superfield $\bG^i(\sigma)$)
which anti-commutes with each others in
the path-integral formalism.
And,
analogous to argument given above (\ref{csresult2}),
the fermion $\eta^i(\sigma)$ are reduced to their zero-modes.
Note that the zero modes satisfy the same
algebra as that for $\frac{\del}{\del (d x^i)}$
or $dx^\alpha$;
$\left\{\frac{\del}{\del(d x^i)},\frac{\del}{\del(d x^j)}\right\}=0$,
which is different from that for
the gamma matrices; $\{ \gamma^i, \gamma^j \} \sim \delta_{ij}$.
This justifies the naive replacement of
$\gamma^i$ to $\frac{\del}{\del (d x^i)}$
given in \cite{AkGeSh}
to show the equivalence between CS-terms
of the BPS D$p$-branes and the non-BPS D$9$-branes.

\subsection{D$p$-branes from unstable D$q$-brane systems}

It is now clear that combining the D-brane ascent
and descent relations described above, we can construct D$p$-branes
from an unstable D$q$-brane system with arbitrary combination of
$p$ and $q$. Let us demonstrate the construction of
a BPS D$p$-brane from non-BPS D$q$-branes.
We consider the non-BPS D$q$-brane
extended along the $x^0, \cdots, x^q$ directions.
Then we construct the D$p$-brane extended along
$x^\alpha$ $(\alpha=0,\ldots, q-m)$
and $x^{\beta}$ $(\beta=q+1,\ldots, q+n)$
 by localizing in the $m$ directions
and extending in the $n$ directions,
where $p=q-m+n$.
The configuration representing the D$p$-brane is
given as a combination of (\ref{Tfluc})--(\ref{Phifluc})
and (\ref{d9});
\beqa
T &=& u \sum_{i=q-m+1}^q
(x^i-\phi^i (x^\alpha,\wh{x}^{\beta})) \otimes \gamma^i
+u \sum_{\beta=q+1}^{q+n}
(\wh{p}_\beta-iA_\beta (x^\alpha,\wh{x}^\beta)) \otimes \gamma^\beta
\label{geneT}
\\
A_\alpha &=& A_\alpha (x^\alpha,\wh{x}^\beta) \otimes 1,
\;\;\;\; A_i=0,
\label{geneA} \\
\Phi^\beta &=& \wh{x}^\beta \otimes 1,
\;\;\;\; \Phi^j=\phi^j (x^\alpha,\wh{x}^\beta) \otimes 1,
\label{genePhi}
\eeqa
where the indices run as
$\alpha=0,\ldots, q-m$,~ $\beta=q+1,\ldots, q+n$,~
$i=q-m+1,\ldots, q$ and $j=q+n+1,\ldots, 9$.
Namely, the indices $(\alpha,i)$ and $(\beta,j)$
correspond to the directions
parallel and perpendicular to the D$q$-branes, respectively,
while $(\alpha,\beta)$ and $(i,j)$ correspond
to directions parallel
and perpendicular to the D$p$-branes, respectively.
We will use the convention of these indices
below in this subsection.
Here, $\gamma^i$ and $\gamma^\beta$ form
the minimally represented $SO(n+m)$ gamma matrices.

Then the boundary state becomes
\beq
\ket{Bq;\pm}_{S_b}
= \int [d \bx^\alpha][d \bx^i]
\,e^{-S_{b}} \,e^{-i\int d\hat\sigma\left(
\bx^\alpha\bP_\alpha+\bx^i\bP_i\right)}
\ket{\bx^\mu=0;\pm},
\eeq
where
\begin{eqnarray}
e^{-S_{b}}&=&\int [d\bx^\beta] [d\bp_\beta]
[d \bG^\beta][d \bG^i]\,
\exp\left\{ \int d\wh\sigma  \left(
\frac{1}{4}\bG^\beta D\bG^\beta +\frac{1}{4}\bG^i D\bG^i +
i\bp_\beta D \bx^\beta
+\bM
\right)
\right\}\nn\\
\end{eqnarray}
with
\begin{eqnarray}
\bM &=&-i\bx^\beta \bP_\beta-i\phi^j(\bx^\alpha, \bx^\beta) \bP_j
-A_\alpha (\bx^\alpha, \bx^\beta) D \bx^\alpha \CR
&& \;\;\;
+ u\left(\bp_\beta-iA_\beta  (\bx^\alpha, \bx^\beta) \right)
\bG^\beta
+ u\left(\bx^i-\phi^i(\bx^\alpha, \bx^\beta)\right)\bG^i.
\end{eqnarray}
Following the previous discussions, we
perform the path-integrations $\int [d \bG^\beta] [d\bp_\beta] $
and $\int [d \bx^i] [d \bG^i] $
which  forces the replacements
$\bG^\beta \ra - \frac{i}{u} D \bx^\beta$ and
$\bG^i \ra \frac{i}{u} \bP^i$, respectively.
Then in the $u \ra \infty$ limit we obtain
\beq
\ket{Bq;\pm}_{S_b}
= \int [d \bx^\alpha]  [d \bx^\beta]
\,e^{-S_{b}}
\ket{\bx^\alpha,\bx^\beta,\bx^i=\bx^j=0;\pm},
\eeq
where
\beqa
S_{b}&=&
\int d\wh\sigma  \left(
A_\alpha (\bx^\alpha, \bx^\beta) D \bx^\alpha
+A_\beta (\bx^\alpha, \bx^\beta) D \bx^\beta
\right.\CR
&&\left.
~~~~ +i\phi^i (\bx^\alpha, \bx^\beta) \bP_i
+i\phi^j (\bx^\alpha, \bx^\beta) \bP_j \right)
.
\eeqa
Therefore we correctly obtain
the boundary state for the BPS D$p$-brane with
the massless fluctuations.

The construction of non-BPS D$p$-brane is obtained by adding
$t(x^\alpha,\wh{x}^\beta)\otimes \gamma_*$ in the tachyon
configuration (\ref{geneT}), where
 $\gamma_*= \prod_{i=q-m+1}^q \gamma^i
\prod_{\beta=q+1}^{q+n}    \gamma^\beta$
up to a phase factor.
If we start with D$q$-\AD$q$ system,
$\gamma^i$ and $\gamma^\beta$ in (\ref{geneT}) need not be hermitian
and they are chosen such that
 $\Gamma^i=\left({~~~\gamma^i\atop\gamma^{i\dag}~~~}\right)$
and
 $\Gamma^\beta=\left({~~~\gamma^\beta\atop\gamma^{\beta\dag}~~~}
\right)$ become $SO(n+m)$ gamma matrices.
In this case, we have two sets of gauge fields
and scalar fields
which are created by D$q$-D$q$ strings and \AD$q$-\AD$q$ strings.
But, we only turn on $A^+$ and $\Phi_+$ in (\ref{bM})
in the same way as given in (\ref{geneA}) and (\ref{genePhi}).
As we can see in (\ref{bM}),
the other ones
$A^-$ and $\Phi_-$ appear
as the higher terms in the gamma matrix expansion
and the they will not contribute in the $u\ra\infty$
limit as explained in section \ref{general}.

\section{Generalization to type I string theory}

In this section, we generalize our argument
in the previous sections to type I string theory.
In the type II string theory, there are basically two
types of unstable D-brane systems, namely the
non-BPS D-branes and the D-brane - anti D-brane systems.
Type I string has eight types of unstable D-brane systems,
which are summarized in table \ref{table}.
They are governed by the real Clifford algebra structure
explained in the next subsection.
We will see that the method using gamma matrix expansion
efficiently works in these unstable D-brane systems.
In this paper, we will only consider turning on the fields
created by D$p$-D$p$ (or D$p$-\AD$p$) strings
and ignore the fields created by the open strings stretched
between the D$p$-brane and one of the background D$9$-branes.
\footnote{See \cite{FGLS} for the study of the type I D-branes
using the boundary state formalism, including
the effect of the states created by the D$p$-D$9$ strings.}

\subsection{Hidden Clifford algebra structure}
\label{hidden}
There is an underlying Clifford algebra
which characterize the structure of Chan-Paton indices for each
unstable D-brane system.
Recall that we used the Pauli matrices $\sigma_1$ and $\sigma_2$
to expand the matrix $\bM$ in (\ref{bM})
for the D$p$-\AD$p$ system in type II string theory.
The Pauli matrices $\sigma_1$ and $\sigma_2$
generate the complex Clifford algebra $\C_2$.
For non-BPS D-branes in type II theory, the matrix $\bM$ is
expanded by $\sigma_1$ which is the generator of $\C_1$.
Here $\C_n$ is the complex Clifford algebra generated by
$n$ elements $e_1,\dots,e_n$ satisfying
\begin{eqnarray}
\{e_i,e_j\}=2\delta_{ij}.
\end{eqnarray}
Note that, as we can see in (\ref{bM}), the tachyon fields are
accompanied by odd numbers of the generators of the Clifford
algebra, while the gauge fields and the scalar fields
are accompanied by even numbers of the generators.

Similarly, in type I string theory, there is a hidden
real Clifford algebra structure in the unstable D-brane
system \cite{AsSuTe2}.
The real Clifford algebra $\C^{r,s}$ is defined as
an algebra over $\R$ generated by $e_i$ ($i=1,\dots,r+s$)
satisfying
\begin{eqnarray}
\begin{array}{ccl}
e_i^2=-1&&(i=1,\dots,r),\\
e_i^2=+1&&(i=r+1,\dots,r+s),\\
e_ie_j+e_je_i=0&&(i\ne j).
\end{array}
\end{eqnarray}
The matrix $\bM$ for the unstable D$p$-brane system with $p=s-r+1$
($\mod 8$) in type I string theory is given by
\begin{eqnarray}
\bM&=&
-\wh A_\alpha(\bX) D\bX^\alpha-i\wh\Phi^i(\bX)\bP_i+\wh T(\bX),
\label{bMtypeI}
\end{eqnarray}
where
\footnote{In this section, every tensor product is defined over $\R$.}
\begin{eqnarray}
\wh A_\alpha(\bX)&=&\sum_{w_n\in\C_{even}^{r,s}}
A_\alpha^{(n)}(\bX)\otimes w_n,\nn\\
\wh\Phi^i(\bX)&=&\sum_{w_n\in\C_{even}^{r,s}}
\Phi^i_{(n)}(\bX)\otimes w_n,\nn\\
\wh T(\bX)&=&\sum_{v_n\in\C_{odd}^{r,s}}T^{(n)}(\bX)
\otimes v_n.
\label{APhiT}
\end{eqnarray}
Here $\C_{even}^{r,s}$ and $\C_{odd}^{r,s}$ denote
the subspaces of $\C^{r,s}$ that consist of elements
which are even and odd under the involution $e_i\ra -e_i$,
respectively, and we sum over the basis $\{w_n\}$
and $\{v_n\}$ of the vector spaces $\C_{even}^{r,s}$ and
$\C_{odd}^{r,s}$, respectively.
The fields $A^{(n)}_\alpha$, $\Phi_{(n)}^i$ and $T^{(n)}$
in (\ref{APhiT}) are real matrices
 (or operator acting on a real Hilbert space).
The generators $e_i$ of the Clifford algebra $\C^{r,s}$
are considered to be fermionic and the choice of
$\C_{even}^{r,s}$ and $\C_{odd}^{r,s}$ in (\ref{APhiT})
is determined so that $\bM$ is a fermionic superfield.
We also impose the hermiticity condition $\bM^\dag = \bM$,
where the hermite conjugate of the generators of the Clifford
algebra are defined as
 $e^*_i=-e_i$ ($i=1,\dots,r$) and $e^*_i=e_i$ ($i=r+1,\dots,r+s$).
In (\ref{bMtypeI}), $D\bX^\alpha$ and $\bP_i$ are
considered to be anti-hermitian, since the time-like and space-like
directions in the open string world-sheet
is interchanged in the closed string
picture. Therefore, the gauge field $\wh A_\alpha$
is anti-hermitian, while $\wh\Phi^i$  and $\wh T$ are hermitian.
It is shown in \cite{AsSuTe2} that the hermiticity condition
of (\ref{bMtypeI}) implies that the gauge group and
the representation of the scalar fields and the tachyon field
with respect to the gauge group are listed in table \ref{table},
which is consistent with the result in \cite{bergman}.
\begin{table}[htb]
\begin{center}
$$
\begin{array}{c|cccc}
\hline\hline
p&\mbox{Gauge}&\mbox{Tachyon}&\mbox{Scalar}&
\mbox{Clifford algebra $\C_{Dp}^{min}$}\\
\hline
-1&U&\asym&\mbox{adj.}&\C^{2,0}\\
0&O&\asym&\sym&\C^{1,0}\\
1&O\times O&(\fnd,\fnd)&(1,\sym),(\sym,1)&\C^{1,1}\\
2&O&\sym&\sym&\C^{0,1}\\
3&U&\sym&\mbox{adj.}&\C^{0,2}\\
4&Sp&\sym&\asym&\C^{0,3}\\
5&Sp\times Sp&(\fnd,\fnd)&(1,\asym),(\asym,1)&\C^{0,4}\\
6&Sp&\asym&\asym&\C^{3,0}\\
7&U&\asym&\mbox{adj.}&\C^{2,0}\\
8&O&\asym&\sym&\C^{1,0}\\
9&O\times O&(\fnd,\fnd)&--&\C^{1,1}\\
\hline
\end{array}
$$
\parbox{75ex}{
\caption{
\small The gauge group and the representation of the tachyon
 and scalar fields on the type I unstable
D$p$-brane system
(non-BPS D$p$-branes for $p=-1,0,2,3,4,6,7,8$
and D$p$-\AD$p$ system for $p=1,5,9$)
 and its underlying Clifford algebra.
Here we listed the
the Clifford algebra which corresponds to
a non BPS D$p$-brane or a pair of D$p$-\AD$p$-branes.}
\label{table}
}
\end{center}
\end{table}

In the table \ref{table}, we listed the minimal choices
of the Clifford algebras which correspond to the unstable
D-brane systems. Other choices of $r$ and $s$ in the
algebra $\C^{r,s}$ with $p=s-r+1$ ($\mod~8$)
will represent $N$ non-BPS D$p$-branes (for $p=-1,0,2,3,4,6,7,8$)
or $N$ pairs of D$p$-\AD$p$-branes (for $p=1,5,9$)
with $N=2^{(r+s-m-n)/2}$
when $\C^{min}_{Dp}=\C^{m,n}$. This is because
there is an isomorphism $\C^{r,s}\simeq M_N(\C^{min}_{Dp})
=M_N(\R)\otimes \C^{min}_{Dp}$ which keeps the involution
\footnote{
In order to keep the involution,
every generator $e_i$ of $\C^{r,s}$ should be mapped to
an element in $M_N(\R)\otimes \C_{odd}^{m,n}$, $i.e.$
the odd elements of $M_N(\C^{m,n})$.}
and we can interpret the $M_{N}(\R)$ part
 as the Chan-Paton
factor for $N$ unstable D$p$-branes.

It is now straightforward to write down the boundary interaction
following the argument around (\ref{super}).
First, we expand the matrix $\bM$ in terms of the generators $e_i$
of the Clifford algebra $\C^{r,s}$ as
\begin{eqnarray}
\bM=\sum_{k=0}^{r+s}\bM^{I_1\cdots I_k}
\otimes e_{I_1\cdots I_k},
\end{eqnarray}
where $e_{I_1\cdots I_k}$ denote the
skew-symmetric product of $e_{I_1},\dots,e_{I_k}$.
Then, our proposal of the boundary interaction
for the type I unstable D$p$-brane is
\begin{eqnarray}
e^{-S_b}&=&\int[d\bG^I]\,\Tr\wh{\rm P}\,
\exp\left\{
\int d\wh\sigma \left(-
\frac{1}{4}\sum_{i=1}^r\bG^i D\bG^i+
\frac{1}{4}\sum_{i=r+1}^{r+s}\bG^i D\bG^i
\right.\right.\nonumber\\
&&\left.\left.
~~~~~~~~~~~~~~+\sum_{k=0}^{r+s}\bM^{I_1\cdots I_k}
\bG^{I_1}\cdots\bG^{I_k}
\right)\right\}.
\label{superI}
\end{eqnarray}

\subsection{D-branes in type I K-matrix theory}
\label{flat}

Let us generalize the construction of D-branes given
in section \ref{DinKmat} to the case of type I string theory.
To be specific, we consider D-branes in the matrix theory
based on non-BPS D-instantons in type I string theory,
which we call type I K-matrix theory.

As listed in table \ref{table}, the underlying Clifford algebra
for the non-BPS D-instanton system is $\C^{2,0}$.
The Clifford algebra $\C^{2,0}$ is faithfully represented
by $e_1=i\sigma_1$ and $e_2=i\sigma_2$.
Then, the tachyon part (the second term)
 of (\ref{bMtypeI}) is of the form
\begin{eqnarray}
\wh T=T^{(1)}\otimes e_1+T^{(2)}\otimes e_2=\mat{,T,-T^*,},
\label{flucF}
\end{eqnarray}
where we have defined $T=T^{(2)}+i T^{(1)}$ and
its complex conjugate $T^*=T^{(2)}-i T^{(1)}$.
In addition, the hermiticity condition
$F^\dag=F$ implies $T=-T^T$.
The scalar part of (\ref{bMtypeI}) is
\begin{eqnarray}
\wh\Phi^\mu=
\Phi^\mu_{(1)}\otimes 1+
\Phi^\mu_{(2)}\otimes e_1e_2=\mat{\Phi^\mu,,,\Phi^{\mu*}},
\end{eqnarray}
where $\Phi^\mu=\Phi^\mu_{(1)}-i\Phi^\mu_{(2)}$
and $\Phi^{\mu*}=\Phi^\mu_{(1)}+i\Phi^\mu_{(2)}$,
and the hermiticity
condition implies $\Phi^{\mu\dag}=\Phi^\mu$.
Here $T$ and $\Phi^\mu$ are the fields on the non-BPS D-instantons,
which are operators acting on a Hilbert space.

The D$p$-brane solution in type I K-matrix theory
is constructed in \cite{AsSuTe2} as
\begin{eqnarray}
\wh T&=&\mat{,T,T^\dag,}
=u\left(\wh\del_0\, e_1
+\sum_{\alpha=1}^p
\wh\del_\alpha \gamma^\alpha_p\, e_2\right),\label{F}\\
\Phi^\alpha&=&\wh x^\alpha~~~(\alpha=0,\dots,p),~~~~~
\Phi^i= 0~~~(i=p+1,\dots,9),
\label{Phi}
\end{eqnarray}
where $\wh\del_\alpha=\del/\del x^\alpha$
 and $\wh x^\alpha$ are the operators
acting on a real Hilbert space $L^2(\R^{p+1})$ and
$\gamma_p^\alpha$ are the $SO(p)$ gamma matrices
represented as real symmetric matrices.
$u$ is a real parameter and $u\ra\infty$
gives the exact solution.
Note that the tachyon operator (\ref{F})
can be rewritten as
\begin{eqnarray}
\wh T&=&
u\,\sum_{\alpha=0}^p
\wh\del_\alpha\wh\Gamma^\alpha,
\label{That}
\end{eqnarray}
where we have set
 $\wh\Gamma^0=e_1$ and $\wh\Gamma^\alpha=\gamma_p^\alpha\,e_2$
~($\alpha=1,\dots,p$). These $\wh\Gamma^\alpha$ satisfies
\begin{eqnarray}
\{\wh\Gamma^\alpha,\wh\Gamma^\beta\}
=-2\delta^{\alpha\beta},~~~(\alpha,\beta=0,\dots,p).
\label{gamma}
\end{eqnarray}
And hence it can be thought of as a natural
analog of the D$p$-brane
solution (\ref{sol1-1}) in type II string theory.
These $\wh\Gamma^\alpha$ are elements of
$M_{n_p}(\C^{2,0})$, where $n_p$ is the size of
the gamma matrices $\gamma^\alpha_p$ listed in table \ref{n_p}.
\begin{table}[htb]
\begin{center}
$$
\begin{array}{c|cccccccccc}
\hline\hline
p&0&1&2&3&4&5&6&7&8&9\\
\hline
n_p&1&1&2&4&8&8&16&16&16&16\\
\hline
\end{array}
$$
\parbox{75ex}{
\caption{
\small
The size of the matrices $\gamma^\alpha$
which is used in the tachyon configuration
(\ref{F}) representing type I D$p$-branes.
}
\label{n_p}
}
\end{center}
\end{table}
We do not have to stick to the explicit form given in (\ref{F}).
$\{\wh\Gamma^\alpha\}$ can be replaced by any set of odd elements
which realize the commutation relation (\ref{gamma})
for the $\C^{p+1,0}$ generators.
Therefore, the D-brane configurations
are obtained by choosing a homomorphism
$\C^{p+1,0}\ra M_{n_p}(\C^{2,0})$.
For $p=1,5$ or $9$, there are two homomorphism which are not
unitary equivalent to each other, which correspond to
a D$p$-brane or an \AD$p$-brane.

In \cite{AsSuTe2}, it has been shown that
this solution exactly gives a correct tension
for a D$p$-brane
(BPS D$p$-brane for $p=1,5,9$ and non-BPS D$p$-brane
for $p=-1,0,2,3,4,6,7,8$) in type I string theory
and the stability analysis of the solutions
consistently reproduces
the spectrum of stable D-branes
expected from K-theory analysis \cite{WittenK}.
Here, we would like to show that
the fluctuations around the D$p$-brane solution
which correspond to the gauge,
scalar and tachyon fields on the D$p$-brane
are precisely what we expect from
the table \ref{table}, by generalizing the
argument given in section \ref{DinKmat}
to type I string theory.
For this, we should reproduce not only the boundary state
representing D$p$-brane but also the correct field content
of the theory, $i.e.$,
real Clifford algebra structure $\C^{min}_{Dp}$.
We will consider
$p\ne 1,5,9$ cases,
though the following argument can also be applied to
a D$p$-\AD$p$ pair with $p=1,5,9$, if
we replace $n_p$ with $2n_p$.
Then, one can show that there is an isomorphism
\begin{eqnarray}
\varphi:\C^{p+1+m,n}\simeq M_{n_p}(\C^{2,0}),
\label{isom}
\end{eqnarray}
where $m$ and $n$ are taken
such that $\C^{min}_{Dp}=\C^{m,n}$.
Therefore, we can realize $\wh\Gamma^\alpha$ in
$M_{n_p}(\C^{2,0})$ by setting
$\wh\Gamma^\alpha\equiv\varphi(e_{\alpha+1})$, ($\alpha=0,\dots,p$).
We extend this definition to span all the generators of $\C^{p+1+m,n}$
as $\wh\Gamma^I\equiv\varphi(e_{I+1})$, ($I=0,\dots,p+m+n$).
We also use the notation
$\wh e_k\equiv\wh\Gamma^{p+k}$, ($k=1,\dots,m+n$) later.
Note that these $\wh e_k$ give a realization of
$\C^{min}_{Dp}=\C^{m,n}$ in $M_{n_p}(\C^{2,0})$.

Using these gamma matrices $\wh\Gamma^I$,
we expand the matrix $\bM$ as
\begin{eqnarray}
\bM&=&
 u\sum_{\alpha=0}^p\wh\del_\alpha\,\wh\Gamma^\alpha
-i\sum_{\alpha=0}^p\wh x^\alpha\bP_\alpha+
\sum_{k=0}^{p+m+n}\delta\bM^{I_1\cdots I_k}
\otimes \wh\Gamma^{I_1\cdots I_k},
\label{Mexp}
\end{eqnarray}
where $\delta\bM^{I_1\cdots I_k}$ represent the fluctuations
around the solution.
The boundary interaction is given as (\ref{superI})
\begin{eqnarray}
e^{-S_b}&=&\int[d\wh\bG^I][d\bx^\alpha]
[d{\bp}_\alpha]
\exp\left\{
\int d\wh\sigma \left(-
\frac{1}{4}\sum_{I=0}^{p+m}\wh\bG^I D\wh\bG^I+
\frac{1}{4}\sum_{I=p+m+1}^{p+m+n}\wh\bG^I D\wh\bG^I
\right.\right.\nonumber\\
&&\left.\left.
~~~~~~~~~~
+\sum_{\alpha=0}^p\left(i\bp_\alpha D\bx^\alpha
+iu\,\bp_\alpha\wh\bG^\alpha
-i\bx^\alpha\bP_\alpha\right)
+\sum_{k=0}^{p+m+n}\delta\bM^{I_1\cdots I_k}
\wh\bG^{I_1}\cdots\wh\bG^{I_k}
\right)\right\}.\nonumber\\
\label{Ibdry}
\end{eqnarray}
Just like (\ref{Dx}),
the path integral with respect to $\bp_\alpha$ implies
\begin{eqnarray}
D\bx^\alpha+u\wh\bG^\alpha=0.
\end{eqnarray}
Then, following the argument leading
(\ref{relevant1}) and (\ref{relevant2}),
we obtain the relevant fluctuations
which correspond to the gauge,
scalar and tachyon fields:
\begin{eqnarray}
\wh T&=&
u\,\sum_{\alpha=0}^p
\left(\wh\del_\alpha+A_\alpha(\wh x)\right)\wh\Gamma^\alpha
+t(\wh x),
\label{Itachyon}
\\
\Phi^i&=& \phi^i(\wh x)~~~(i=p+1,\dots,9).
\label{Iscalar}
\end{eqnarray}
Here $A_\alpha(\wh x)$, $t(\wh x)$ and $\phi^i(\wh x)$
are matrices which can be expanded by
$\wh e_k=\wh\Gamma^{p+k}$, ($k=1,\dots,m+n$).

Recall that $\wh e_k$ can be thought of as
the generators of $\C^{min}_{Dp}$.
Since the operator $\wh T$ is odd under the
involution $e_i\ra-e_i$,
$A_\alpha(\wh x)$ and $t(\wh x)$ should be
even and odd elements of $\C^{min}_{Dp}$, respectively.
Furthermore, from the hermiticity condition
$\wh T^\dag=\wh T$,
$A_\alpha(\wh x)$ and $t(\wh x)$ should be anti-hermite
and hermite operators, respectively. Similarly,
$\phi^i(\wh x)$ are hermite operators
and even elements of $\C^{min}_{Dp}$.
These properties are exactly what we have imposed
for the fields on the type I D$p$-brane
to obtain the table \ref{table}.
Inserting (\ref{Itachyon}) and (\ref{Iscalar})
in (\ref{Ibdry})
 and performing the path
integral with respect to $\bp_\alpha$,
we correctly obtain the boundary state
for D$p$-brane with boundary interaction
 (\ref{bMtypeI}) in the limit $u\ra\infty$.

\subsection{CS-term, real superconnections and the index theorem}
\label{typeICSind}
It is interesting to apply the arguments given in
section \ref{cs-term} to the type I case.
The CS-term for unstable D$p$-branes in type I string theory
can also be written like (\ref{DpCS}),
where (\ref{J}) is now given as
\begin{eqnarray}
J(x^\alpha,k_i,\psi_1^\alpha,\psi_2^i)
=\Str\left( e^{-ik_i\wh \Phi^i+2\pi\wh\cF}\right),
\label{JtypeI}
\end{eqnarray}
\begin{eqnarray}
\wh\cF=
\half\wh F_{\alpha\beta}\psi_1^\alpha\psi_1^\beta-\wh T^2
+\frac{1}{8\pi^2}[\wh\Phi^i,\wh\Phi^j]\psi_2^i\psi_2^j
-\frac{1}{2\pi}\psi_1^\alpha \psi_2^i D_\alpha \wh\Phi^i
+i\psi_1^\alpha D_\alpha \wh T-\frac{i}{2\pi}\psi_2^i[\wh\Phi^i,\wh T],
\nn\\
\label{typeIF}
\end{eqnarray}
using the notation in (\ref{APhiT}) and
$D_\alpha=\del_\alpha+\wh A_\alpha$.
 $\Str$ in (\ref{JtypeI}) denotes the
supertrace which is proportional to
the trace of the coefficient of $e_1e_2\cdots e_{r+s}$.
$\wh\cF$ can also be written as
\begin{eqnarray}
\wh\cF=-Z^2,~~~
Z=-i\psi^\alpha_1 (\del_\alpha+\wh A_\alpha)+\frac{i}{2\pi}
\psi_2^i\wh\Phi^i+\wh T.
\end{eqnarray}

This $\wh\cF$ is a analog of the
field strength of the superconnection.
Recall that there are two types of (complex)
superconnections \cite{Qu}
based on complex Clifford algebras,
called even and odd superconnections,
and they naturally
appear in the CS-terms of D-brane - anti D-brane
systems and non-BPS D-branes \cite{KrLa,TaTeUe,KeWi},
as we have seen in section \ref{genCS}.
Generalizing the definition of the superconnection
using real Clifford algebra,
we can define eight types of real superconnections
associated with the eight types of
unstable D-brane systems listed in table \ref{table}.
When we turn off the scalar fields $\wh \Phi^i$,
the CS-term is simplified as \cite{AsSuTe2}
\begin{eqnarray}
S_{CS}^{Dp}&=&\mu_p\int_{Dp} C\wedge \Str
e^{2\pi\wh\cF}
,\\
\wh\cF&=&-\wh T^2+iD_\alpha \wh T\, dx^\alpha
+\half\wh F_{\alpha\beta}\, dx^\alpha\wedge dx^\beta.
\end{eqnarray}

We will now show that only $n$-form part
of $e^{\wh\cF}$  with
$n\equiv p-1$ ($\mod 4$) will survive
after taking the supertrace.
This is consistent with the fact that the
R-R $p$-form fields $C_p$ only exist for $p=2,6,(10)$
in type I string theory.
\footnote{The coupling of R-R $10$-form
is canceled by the contribution
from the $O9^-$-plane.}
First, note that
the $n$-form part of $e^{\wh\cF}$
which contributes in the supertrace is of the form
\begin{eqnarray}
H\equiv i^n G\,e_1e_2\cdots e_{r+s}\,dx^1\cdots dx^n,
\end{eqnarray}
where $G$ is a real matrix. Since $\wh\cF$ is required to be hermite
and bosonic, $H$ is also hermite and $r+s+n\equiv 2k$ is an even number.
Here $dx^\alpha$ is treated as a fermion since it came from
$\psi_1^\alpha$ in (\ref{typeIF}).
Then, hermite conjugate of $H$ is given by
\begin{eqnarray}
H^\dag
&=&(-1)^{\half(r+s+n)(r+s+n-1)-r-n}\,i^nG^T
\,e_1e_2\cdots e_{r+s}\,dx^1\cdots dx^n\\
&=&(-1)^{k-r-n}\,i^nG^T
\,e_1e_2\cdots e_{r+s}\,dx^1\cdots dx^n.
\end{eqnarray}
Therefore the hermiticity of $H$ implies
\begin{eqnarray}
G=(-1)^{k-r-n}G^T.
\end{eqnarray}
Since the supertrace of $e^{\wh\cF}$ is
proprotional to the trace over $G$,
$k-r-n$ should be an even number to obtain a non-zero contribution.
Then we obtain $n\equiv s-r\equiv p-1$ ($\mod 4$).

It is straightforward to generalize the argument given
in section \ref{CSascent} and \ref{index} to
show the descent/ascent relations of the CS-terms
for the type I D-branes
and to obtain index theorems for the gauge theories
listed in the table \ref{table}.

As an example,
let us start with D$1$-\AD$1$ system and
consider the homogeneous configuration along the world-volume
of the D$1$-branes. The D$1$-brane charge of the system
can be read from the coefficient of R-R 2-form field
$C_2$ in the CS-term and it is
 again given by the index of the tachyon operator
on the D$1$-\AD$1$ system.
When we construct D$p$-branes in this system,
the tachyon operator becomes a Dirac operator
for the gauge theory on the D$p$-brane.
On the other hand, from the viewpoint
of the D$p$-brane world-volume gauge theory,
the D$1$-brane charge is given as
the integral of $\Str e^{\wh\cF}$ along the direction
transverse to the D$1$-brane in the D$p$-brane
world-volume.
Comparing the two descriptions, we conclude that
the index of the Dirac operator is given by
$\int\Str e^{\wh\cF}$.

We can also apply this strategy for the
$\Z_2$ charged stable non-BPS D-branes
in type I string theory found in \cite{Sen,WittenK}.
As shown in \cite{WittenK}, the D-brane charge
is classified by the real K-theory.
More specifically, flat D$p$-branes in the flat space-time
 are classified by
real K-theory group $KO(\R^{9-p})$, which is
$\Z$ for $p=1,5,9$ and $\Z_2$ for $p=-1,0,7,8$.
These $\Z_2$ charges for the D$(-1),0,7,8$-branes
cannot be written as the integral of a differential form
unlike the $\Z$ charge of D$1,5,9$-branes.
However, it is still possible to interpret the charge
in terms of
the configurations of tachyon and gauge fields on
D$9$-\AD$9$ system \cite{WittenK}
or other higher dimensional unstable D-brane systems
\cite{bergman} in type I string theory.
On the other hand, as explained in \cite{AsSuTe2},
the $\Z_2$ charge of unstable D-branes are also
obtained as an index of the tachyon operators.
In fact, as we can see from the table \ref{table},
 the tachyon of the D$(-1),0,7,8$-branes
transform as the anti-symmetric tensor representation
of the gauge group,
\footnote{The tachyon on the D$6$-brane is also
anti-symmetric. But, the anti-symmetric tensor
representation of $Sp$ group does not carry the index.}
and hence the $\mod 2$ index, ${\rm Index}\,T\equiv
\dim{\rm Ker}\, T$ ($\mod 2$), is invariant
under small perturbations of the operator $T$.
Again, comparing the two descriptions, we can
obtain the relations between the $\mod 2$ indices of Dirac operators
and the field configurations of the gauge theory
on the higher dimensional unstable D-brane systems.

We will not study these issues further in this paper.
It would be very interesting to explore
the descent/ascent
relations of various topological invariants
more extensively using this approach.

\section{Conclusions and Discussions}

In this paper, we gave a simple derivation
of D-brane descent/ascent
relations. Taking the fluctuations around a D-brane
solutions into consideration, we
 correctly reproduced the D-brane boundary state
with boundary interaction.
We found that the superfield formulation was
very useful for the analysis.
We analyzed type II as well as type I string theory.
The Clifford algebra structure
is efficiently used in the analysis of type I
D-branes.

Since we have seen how to realize the D-brane action from
K-matrix theory, we can in principle analyze anything
which have been done using D-brane action in the framework of
K-matrix theory.
For example, since we can also obtain non-BPS D-brane action
from K-matrix theory, it is possible to repeat the
analysis given in \cite{SuTe,Mi} for the
rolling of the tachyon field on the non-BPS D-brane using
K-matrix theory.

It is important to note that
K-matrix theory makes it possible to construct
any configurations of D-branes with various
dimensions in a single set-up.
We can also handle creation and annihilation of
unstable D-branes which plays an important role
in string theory.
We haven't analyzed fundamental strings in K-matrix theory.
But, since the fundamental strings
can be realized as electric flux tubes in unstable
D-brane system \cite{Yi,BeHoYi,HaKrLaMa},
it should be possible to realize them
as such configurations in K-matrix theory.
Therefore, it is reasonable to expect that the
theory contains the degrees of freedom of
fundamental strings as well. It would be interesting to
investigate this further.

Another interesting question is
how supersymmetry of the BPS D-branes constructed
from the non BPS D-branes is realized.
There should be a nonlinear supersymmetry
on the unstable D-brane system and
its restoration is expected to explain the BPS property
\cite{TeUe,SenN,Yo}.
It may be interesting to study it using the boundary states.

The field variables in K-matrix theory
 consists of operators acting on
an infinite dimensional Hilbert space.
If we regularize the theory by replacing the operators
with finite size matrices, it provides a kind
of lattice regularization of gauge theory on D-branes.
It would be interesting to see if this approach could
give a constructive definition of higher dimensional
gauge theory or string theory.

As for the action of K-matrix theory, we adopted BSFT
action in this paper. It is an excellent approach
for the consideration of tachyon condensation.
 However, there are
some problems concerned with massive modes \cite{LiWi}.
In order to discuss quantum effects, it is inevitable
to include these massive modes.
We leave these issues for the future investigation.

\section*{Acknowledgments}
\vskip2mm

We would like to thank K. Hashimoto for useful discussions.
S.T. is grateful to the Michigan center for theoretical physics
and the Harvard high energy theory group for hospitality.
S.S. also thanks high energy theory groups at
Yukawa institute for theoretical
physics and Seoul national university
for hospitality and especially grateful
to Soo-Jong Rey for various enjoyable discussions.
The work of A.T. is supported in part by
JSPS Research Fellowships for Young Scientists.
The work of S.S. is supported in part by
Danish Natural Science Research Council.

\appendix
\setcounter{equation}{0}

\section{Operator vs. path integral}
\label{Oppath}

In this appendix,
we will show the equivalence between operator and
path integral formulations
in the superfield formalism used in section \ref{DinKmat}.
We will compare the following two objects.
\begin{eqnarray}
Z_1&=&\Tr\wh{\rm P}\, e^{\int d\hat\sigma \bM(\hat x,\hat p)}
\label{Z-1}\\
Z_2&=&\int[d\bx][d\bp]\,
\exp\left\{
\int d\wh\sigma\left(\,
i\bp D\bx+\bM(\bx,\bp)
\,\right)\right\}.
\label{Z-2}
\end{eqnarray}

As explained in (\ref{SP}), the supersymmetric path ordered
exponential in $Z_1$ is defined as
\begin{eqnarray}
Z_1&=&
\sum_{n=0}^\infty
\int d\wh\sigma_1\cdots d\wh\sigma_n
\Theta(\wh\sigma_{12})\Theta(\wh\sigma_{23})\cdots
\Theta(\wh\sigma_{n-1\,n})\,\nn\\
&&~~~\times(-1)^{\frac{n(n-1)}{2}}
\Tr\left\{
\,\bM(\hat x,\hat p)_{\hat\sigma_1}
\cdots\bM(\hat x,\hat p)_{\hat\sigma_n}
\right\},\nn\\
\end{eqnarray}
where $\bM(\hat x,\hat p)_{\hat\sigma_i}$ denotes
 $\bM(\hat x,\hat p)$ evaluated at $\wh\sigma=\wh\sigma_i$.
If we rewrite this formula using the standard
notations for non-commutative field theory
\cite{AIIKKT}, we obtain
\begin{eqnarray}
Z_1&=&
\sum_{n=0}^\infty
\int d\wh\sigma_1\cdots d\wh\sigma_n
\Theta(\wh\sigma_{12})\Theta(\wh\sigma_{23})\cdots
\Theta(\wh\sigma_{n-1\,n}) \nn\\
&&~~~\times (-1)^{\frac{n(n-1)}{2}}
\frac{1}{2\pi}\int dx dp
\,\bM(x,p)_{\hat\sigma_1}*
\cdots *\bM(x,p)_{\hat\sigma_n},
\label{noncomm}
\end{eqnarray}
where $*$ is the star product defined as
\begin{eqnarray}
A(x,p)* B(x,p)=\left.
 e^{\frac{i}{2}
 \left(\frac{\del}{\del x_1}\frac{\del}{\del p_2}
 -\frac{\del}{\del p_1}\frac{\del}{\del x_2}\right)}
A(x_1,p_1) B(x_2,p_2)\right|_{x=x_1=x_2\atop p=p_1=p_2}
\end{eqnarray}

To compare $Z_1$ and $Z_2$ order by order, we also expand $Z_2$
as
\begin{eqnarray}
Z_2=\sum_{n=0}^\infty
\frac{1}{n!}
\int d\wh\sigma_1\cdots d\wh\sigma_n
(-1)^{\frac{n(n-1)}{2}}
\int[d\bx][d\bp]\,e^{i\int d\hat\sigma\,\bp D\bx}
\,\bM(\bx,\bp)_{\hat\sigma_1}
\cdots \bM(\bx,\bp)_{\hat\sigma_n}
\label{z2exp}
\end{eqnarray}
The path integral can be evaluated
by the usual technique of perturbation theory as
\begin{eqnarray}
&&\int[d\bx][d\bp]\,e^{i\int d\hat\sigma\,\bp D\bx}
\,\bM(\bx,\bp)_{\hat\sigma_1}
\cdots \bM(\bx,\bp)_{\hat\sigma_n}
\label{xp0}
\\
&=& \frac{1}{2\pi}\int dx_0dp_0
\left.\,e^{\frac{\delta}{\delta \bp}\cdot\Theta\frac{\delta}{\delta \bx}}
\,\bM(\bx,\bp)_{\hat\sigma_1}
\cdots \bM(\bx,\bp)_{\hat\sigma_n}
\right|_{\bx(\hat\sigma)=x_0\atop\bp(\hat\sigma)=p_0},
\label{xp}
\end{eqnarray}
where we have defined
\begin{eqnarray}
\frac{\delta}{\delta \bp}\cdot\Theta\frac{\delta}{\delta \bx}
&=&
\frac{i}{4}
\int d\wh\sigma_a\int d\wh\sigma_b
\,{\rm sign}(\wh\sigma_{ab})\left(
\frac{\delta}{\delta \bx(\wh\sigma_a)}
\frac{\delta}{\delta \bp(\wh\sigma_b)}
-\frac{\delta}{\delta \bp(\wh\sigma_a)}
\frac{\delta}{\delta \bx(\wh\sigma_b)}
\right)
\CR
&=& \frac{i}{2}
\int d\wh\sigma_a\int d\wh\sigma_b
\,\Theta(\wh\sigma_{ab})\left(
\frac{\delta}{\delta \bx(\wh\sigma_a)}
\frac{\delta}{\delta \bp(\wh\sigma_b)}
-\frac{\delta}{\delta \bp(\wh\sigma_a)}
\frac{\delta}{\delta \bx(\wh\sigma_b)}
\right),
\label{propa}
\end{eqnarray}
where
${\rm sign}(\wh\sigma_{ab})=
\Theta(\wh\sigma_{ab})-\Theta(\wh\sigma_{ba})$
is the sign function.
The normalization factor $1/2\pi$ in (\ref{xp})
comes from the path integral measure which will be fixed
in Appendix \ref{Normalization}.
(\ref{xp}) is computed by contracting $\bx$ and $\bp$
with the propagator. Here, the choice of the
constant term in the propagator does not affect the
final result and the expression (\ref{xp})
has no ambiguity.
We will use
$\Theta(\wh\sigma_{ab})$ as the propagator
which turns out to be very convenient
in order to show $Z_1=Z_2$.

When we compute (\ref{xp}) by contracting $\bx$ and $\bp$
with the propagators, we encounter products
of $\Theta(\wh\sigma_{ab})$'s.
However, since  $\Theta(\wh\sigma_{ab})$ is defined as
\begin{eqnarray}
\Theta(\wh\sigma_{ab})&=&\Theta(\sigma_a-\sigma_b-\theta_a\theta_b)\nn\\
&=&\Theta(\sigma_a-\sigma_b)
-\theta_a\theta_b\delta(\sigma_a-\sigma_b),
\end{eqnarray}
we have to deal with the products of the delta function and
step functions, which are not well-defined. In order to fix this
ambiguity, we modify the step function
$\Theta(\wh\sigma_{ij})$
in ({\ref{propa}) as
\begin{eqnarray}
\Theta_\epsilon(x)=\left\{
\begin{array}{cc}
0&x<-\epsilon\\
\half+\frac{x}{2\epsilon}&-\epsilon \le x \le\epsilon\\
1&\epsilon< x,
\end{array}
\right.
\end{eqnarray}
and take the limit $\epsilon\ra +0$ afterward.
Then, the product of a step function and a delta function
is understood as
$\Theta(\sigma_a-\sigma_b)\delta(\sigma_a-\sigma_b)=
\half\delta(\sigma_a-\sigma_b)$. It is easy to check
the identities
\begin{eqnarray}
\Theta(\wh\sigma_{ab})^2&=&\Theta(\wh\sigma_{ab}),
\label{identity1}\\
\Theta(\wh\sigma_{ab})\Theta(\wh\sigma_{ba})&=&0,\\
\Theta(\wh\sigma_{ab})\Theta(\wh\sigma_{bc})
\Theta(\wh\sigma_{ac})
&=&\Theta(\wh\sigma_{ab})\Theta(\wh\sigma_{bc}),
\label{identity2}
\end{eqnarray}
$etc.$, which are satisfied
up to some functions with measure zero support.

Note also that we have
\begin{eqnarray}
\sum_{s\in S_n} \Theta(\wh\sigma_{s(1)s(2)})\cdots
 \Theta(\wh\sigma_{s(n-1)s(n)})=1,
\label{perm}
\end{eqnarray}
where $S_n$ is the permutation group of $n$ indices.
{}From (\ref{perm}), we can replace
the $\wh\sigma$ integral in (\ref{z2exp}) as
\begin{eqnarray}
\frac{1}{n!}\int d\wh\sigma_1\cdots d\wh\sigma_n
&=&\frac{1}{n!}
\sum_{s\in S_n}
\int d\wh\sigma_1\cdots d\wh\sigma_n
\Theta(\wh\sigma_{s(1)s(2)})\cdots
 \Theta(\wh\sigma_{s(n-1)s(n)})\\
&\ra&
\int d\wh\sigma_1\cdots d\wh\sigma_n
 \Theta(\wh\sigma_{12})\cdots \Theta(\wh\sigma_{n-1\,n}),
\label{perm1}
\end{eqnarray}
using the permutation invariance of
 $d\wh\sigma_1\bM_{\hat\sigma_1}\cdots
 d\wh\sigma_n\bM_{\hat\sigma_n}$
 in (\ref{z2exp}).
Then, using the identities among the propagators
such as (\ref{identity1})--(\ref{identity2}),
the contractions with the propagators $\Theta(\wh\sigma_{ij})$
in (\ref{xp}) can be
reduced to the star products.
\begin{eqnarray}
&&
\Theta(\wh\sigma_{12})\cdots\Theta(\wh\sigma_{n-1\,n})
\int dx_0dp_0
\left.\,e^{\frac{\delta}{\delta \bp}\cdot\Theta\frac{\delta}{\delta \bx}}
\,\bM(\bx,\bp)_{\hat\sigma_1}
\cdots \bM(\bx,\bp)_{\hat\sigma_n}
\right|_{\bx(\hat\sigma)=x_0\atop \bp(\hat\sigma)=p_0}\nn\\
&=&
\Theta(\wh\sigma_{12})\cdots\Theta(\wh\sigma_{n-1\,n})
\int dx dp\, \bM(x,p)_{\sigma_1}*\cdots *\bM(x,p)_{\sigma_n}
\end{eqnarray}
Here we have used the fact that there is no contraction between
the $\bx$ and $\bp$ in the same $\bM(\bx,\bp)_{\hat\sigma_i}$ factor.
Inserting this into (\ref{z2exp}) and comparing it with
(\ref{noncomm}),
we obtain the desired equivalence $Z_1=Z_2$.

We can easily
generalize this argument to the analogous statement
for fermionic boundary super fields.
Let us now show $Z_1=Z_2$, where
\begin{eqnarray}
Z_1&=&\kappa\Tr\wh{\rm P}\, e^{\int\! d\hat\sigma \bM(\Gamma)} \;\;
{\rm (NSNS)}, \;\;\;\;\;\;
Z_1=\Str \wh{\rm P}\, e^{\int\! d\hat\sigma \bM(\Gamma)} \;\;
{\rm (RR)},
\\
Z_2&=&\int[d\bG]\,
\exp\left\{
\int d\wh\sigma\left(\,
\frac{1}{4} \bG D\bG+\bM(\bG)
\,\right)\right\},
\label{Z2}
\end{eqnarray}
where $\Gamma=(\Gamma^1,\dots,\Gamma^n)$ are
gamma matrices
and $\bG=(\bG^1,\dots,\bG^n)$ are
corresponding superfields. The normalization
constant $\kappa$ is $\kappa=1$ or $\kappa=1/\sqrt{2}$
when $n$ is even or odd, respectively.
This statement was proved in \cite{Marcus} using
the point splitting regularization.
Our method is useful because of its manifestation of supersymmetry,
although result should be regularization independent.

First we introduce fermionic $*$-product between
two polynomials of classical fermion fields $\eta^i$ as
\beq
A(\eta)* B(\eta) = \left.
e^{
-\frac{\del}{\del \eta_1^i} \frac{\del}{\del \eta_2^i}}
A(\eta_1) B(\eta_2) \right|_{\eta^i=\eta_1^i=\eta_2^i}.
\eeq
This corresponds to the algebra among
 anti-symmetrized polynomials of
the gamma matrix $\Gamma^i$.
Actually, we can easily see that
$A(\eta) * B(\eta)$
corresponds to $A(\Gamma) B(\Gamma)$
as in the bosonic case, where
contractions of the Gamma matrices $(\Gamma^i)^2=1$ is represented
by the fermionic $*$-product.
It is also seen that
$\kappa\Tr(A(\Gamma))$ and
$\Str(A(\Gamma))=\kappa\Tr\left((-i)^{n/2}\prod_{i=1}^n \Gamma^i
\, A(\Gamma)\right)$
correspond to $2^{n/2} A(\eta)|_{\eta^i=0}$
and $(-2i)^{n/2}\int d\eta^1 \cdots d\eta^n A(\eta)$,
respectively.

The propagator for $\bG$
in the path integral (\ref{Z2})
is given by the sign function and we have
\begin{eqnarray}
&&\int[d\bG]\,e^{ \frac{1}{4} \int d\hat\sigma\,\bG D\bG}
\,\bM(\bG)_{\hat\sigma_1}
\cdots \bM(\bG)_{\hat\sigma_n}\\
&=&(-2i)^{n/2}\int \prod_i d\eta_0^i\,
\left.\,e^{\frac{\delta}{\delta \bG}\cdot\Theta\frac{\delta}{\delta \bG}}
\,\bM(\bG)_{\hat\sigma_1}
\cdots \bM(\bG)_{\hat\sigma_n}
\right|_{\bG^i(\hat\sigma)=\eta_0^i},
\label{eta}
\end{eqnarray}
for RR-sector and
$2^{n/2}
\left.\,e^{\frac{\delta}{\delta \bG}\cdot\Theta\frac{\delta}{\delta \bG}}
\,\bM(\bG)_{\hat\sigma_1}
\cdots \bM(\bG)_{\hat\sigma_n}
\right|_{\bG^i(\hat\sigma)=0}$ for NSNS-sector.
Here we have defined
\begin{eqnarray}
\frac{\delta}{\delta \bG}\cdot\Theta\frac{\delta}{\delta \bG}
&=&-
\int d\wh\sigma_a\int d\wh\sigma_b
\,\Theta(\wh\sigma_{ab} )\left(
\frac{\delta}{\delta \bG(\wh\sigma_a)}
\frac{\delta}{\delta \bG(\wh\sigma_b)}
\right).
\end{eqnarray}
{}From these we obtain $Z_1=Z_2$ repeating the argument
for the bosonic case.

Combining these results,
we can directly show
the equivalence between path integral and operator formulation
of the boundary interaction used in the paper.

\section{Normalization}
\label{Normalization}

In order to fix the normalization, we have to determine the path
integral measure. We recall the equivalence between the
operator formulation and the path integral formulation
of a free particle;
\begin{eqnarray}
\Tr \left(e^{-i\int\! d\sigma \frac{\hat p^2}{2m}}\right)
=\int [dx]\, e^{i\int\! d\sigma \frac{m}{2}\dot x^2}.
\label{particle}
\end{eqnarray}
We use this equation to define the path integral
measure for the bosonic variables.
The left hand side of (\ref{particle}) can be evaluated as
\begin{eqnarray}
\Tr \left(e^{-2\pi i \frac{\hat p^2}{2m}}\right)
=\int dp \bra{p}e^{-2\pi i \frac{\hat p^2}{2m}}\ket{p}
=\int dp\, e^{-2\pi i \frac{p^2}{2m}} \bracket{p}{p}
=\frac{\sqrt{-im}}{2\pi}\int dx.
\label{opresult}
\end{eqnarray}
On the other hand, the right hand side of (\ref{particle}) will become
\begin{eqnarray}
\int [dx]\, e^{i\int\! d\sigma \frac{m}{2}\dot x^2}
&=&K\int dx_0\prod_{n=1}^\infty dx_n dx_{-n}
\,e^{2\pi i m \sum_{n=1}^\infty n x_{-n}x_n}\\
&=&K\int dx_0\prod_{n=1}^\infty\frac{\pi}{-2\pi i mn}
=K\sqrt{\frac{-im}{\pi}}\int dx_0,
\label{pathresult}
\end{eqnarray}
where $K$ is a normalization constant.
Here we have used the following formulae obtained
by zeta function regularization;
\begin{eqnarray}
\prod_{n=1}^\infty A=A^{-1/2},~~
\prod_{n=1}^\infty \frac{1}{n}=\frac{1}{\sqrt{2\pi}}.
\end{eqnarray}
In order to equate (\ref{opresult}) and (\ref{pathresult}),
we should set
\begin{eqnarray}
K=\frac{1}{2\sqrt{\pi}}.
\end{eqnarray}
With this normalization, we can also show
\begin{eqnarray}
\int[dp]\,e^{-i\int\! d\sigma\,\frac{1}{2m} p^2}
&=&K\int dp_0\,e^{-\frac{\pi i}{m}p_0^2}
\int\prod_{n=1}^\infty dp_ndp_{-n}
e^{-\frac{2\pi i}{m}\sum_{n=1}^\infty \frac{1}{n}p_{-n}p_n}\nn\\
&=&K\sqrt{-im}\prod_{n=1}^\infty\frac{\pi m n}{2\pi i}
=K 2\sqrt{\pi}\nn\\
&=& 1,
\label{gausspath}
\end{eqnarray}
which implies, as usual,
\begin{eqnarray}
\int [dx]\, e^{i\int\! d\sigma \frac{m}{2}\dot x^2}
=\int[dx][dp]\,e^{i\int\! d\sigma\left(p\dot x-\frac{1}{2m} p^2\right)}.
\label{phasepath}
\end{eqnarray}
If we only perform the integral with respect to non-zero modes in
the right hand side of (\ref{phasepath}), we obtain
\begin{eqnarray}
&&\int[dx][dp]\,e^{i\int\! d\sigma\left(p\dot x-\frac{1}{2m} p^2\right)}\\
&=&K^2\int dx_0dp_0 \,e^{-\frac{\pi i}{m}p_0^2}
\int\prod_{n=1}^\infty dx_n dx_{-n} dp_n dp_{-n}
\,e^{2\pi i\sum_{n=1}^\infty
(ip_nx_{-n}-ip_{-n}x_n-\frac{1}{mn}p_{-n}p_n)}\nn\\ \\
&=&K^2\prod_{n=1}^{\infty}
\left(\frac{\pi mn}{2\pi i}\frac{\pi}{-2\pi i mn}
\right)\int dx_0dp_0 \,e^{-\frac{\pi i}{m}p_0^2}\\
&=&\frac{1}{2\pi}\int dx_0dp_0 \,e^{-\frac{\pi i}{m}p_0^2}.
\label{2pi}
\end{eqnarray}
This normalization factor $1/2\pi$ in (\ref{2pi}) is
the origin of the factor $1/2\pi$ that
we encountered in (\ref{xp}).

For the fermions, we can use the
path integral representation of matrices
explained around (\ref{bint}). Recall that
2$\times$2 matrices can be
expanded by Pauli matrices $\sigma^1$ and $\sigma^2$,
and we can represent them by fermions
$\eta^1$ and $\eta^2$ in the path integral. Then,
for the NS-NS sector, we have
\begin{eqnarray}
2&=&\Tr\left(\,1~~~\,\atop\,~~~1\,\right)=
\int[d\eta^1][d\eta^2]\,
 e^{\int\! d\sigma \frac{1}{4}
 \left(\dot\eta^1\eta^1+\dot\eta^2\eta^2\right)}\\
&=&
\left(K_\eta
\int\prod_{r=1/2}^\infty d\eta^1_{r}d\eta^1_{-r}
 e^{\sum_{r=1/2}^\infty
i\pi r \eta^1_{-r}\eta^1_{r}}\right)^2\\
&=&
\left(K_\eta\prod_{r=1/2}^\infty (i\pi r)
\right)^2=2 K_\eta^2.
\end{eqnarray}
Therefore the normalization constant is $K_\eta=1$.
Here we have used the following
the zeta function regularization formulae;
\begin{eqnarray}
\prod_{r=1/2,3/2,\cdots}^\infty A=1,~~
\prod_{r=1/2,3/2,\cdots}^\infty r=\sqrt{2}.
\end{eqnarray}
In particular, we obtain
\begin{eqnarray}
\int[d\eta]\,e^{\int\! d\sigma\frac{1}{4}\dot\eta\eta}=\sqrt{2}
\label{sqrt2}
\end{eqnarray}
for the NS-NS sector boundary fermion.
The analogous formula in superfield
formulation is also useful;
\begin{eqnarray}
\int[d\bG]\,e^{\int\! d\hat\sigma\frac{1}{4}\bG D\bG}
=\int[d\eta][dF]\,e^{\int\! d\sigma\frac{1}{4}
\left(\dot\eta\eta+F^2\right)}=\sqrt{2},
\end{eqnarray}
where we have used the analytic continuation of the
formula (\ref{gausspath}) for the integration
with respect to the auxiliary field $F$.

Similarly, the normalization constant for the R-R sector is
determined by
\begin{eqnarray}
2&=&\Str \sigma^3 =
\int[d\eta^1][d\eta^2]\,i\eta_0^2\eta_0^1
\, e^{\int\! d\sigma \frac{1}{4}
 \left(\dot\eta^1\eta^1+\dot\eta^2\eta^2\right)}\\
&=&i
\left(K_\eta
\int d\eta_0^1\,\eta_0^1\,\int
\prod_{n=1}^\infty d\eta^1_{n}d\eta^1_{-n}
 e^{\sum_{n=1}^\infty
i\pi n \eta^1_{-n}\eta^1_n}\right)^2\\
&=&i
\left(K_\eta\prod_{n=1}^\infty (i\pi n)
\right)^2=2 K_\eta^2,
\end{eqnarray}
which again implies $K_\eta=1$. Thus we obtain
\begin{eqnarray}
\int[d\eta]\,\eta_0\,e^{\int\! d\sigma\frac{1}{4}\dot\eta\eta}
&=&\sqrt{-2\,i}
\end{eqnarray}
for the R-R sector boundary fermion.

Using these definition of the path integral measures,
we can fix the overall normalization of the
D$p$-brane action (\ref{BSFT}) as
\begin{eqnarray}
S^{Dp}=\frac{2\pi}{g_s}\int[d\bx^\alpha]
\bra{0} e^{-S_b}\ket{\,\bx^\alpha,\bx^i=0;+}_{\rm NS}.
\label{Snorm}
\end{eqnarray}
In fact, the tension for a BPS D$p$-brane is correctly
obtained by turning off the fields on the D$p$-brane;
\begin{eqnarray}
&&\frac{2\pi}{g_s}\int[d\bx^\alpha]\bracket{0}{\,\bx^\alpha;+}_{\rm NS}\\
&=&\frac{2\pi}{g_s}
\prod_{\alpha=0}^p\left(K\int
dx_0^\alpha\prod_{n=1}^\infty dx^\alpha_n dx^\alpha_{-n}
\prod_{r=1/2}^\infty d\psi^\alpha_r d\psi^\alpha_{-r}
\,e^{-\half\sum_{n=1}^\infty x_{-n}^\alpha x_{n}^\alpha
-\half\sum_{r=1/2}^\infty\psi_{-r}^\alpha\psi_{r}^\alpha}
\right)\nn\\ \\
&=&\frac{2\pi}{g_s}\left(\frac{1}{2\sqrt{2}\pi}\right)^{p+1}\int d^{p+1}x_0
={\cal T}_p\int d^{p+1}x_0,
\end{eqnarray}
where
\begin{eqnarray}
{\cal T}_{p}=\frac{1}{(2\pi)^{p}(\sqrt{\alpha'})^{p+1}g_s}
\end{eqnarray}
is the tension for the D$p$-brane with the convention
$\alpha'=2$.
For the non-BPS D$p$-branes, we have
an additional $\sqrt{2}$ factor which comes
 from the contribution of
the boundary fermion (\ref{sqrt2}).
It is important to note that the normalization
constant $2\pi/g_s$ in (\ref{Snorm}) is independent of $p$.
Therefore, once we obtain the descent/ascent relations
among the boundary states $\ket{Bp;+}_{S_b}$, we can exactly
reproduce the D$p$-brane action including the tension.

The coefficient for the CS-term $\mu_p$ in (\ref{DpCS}) is
similarly determined as
\begin{eqnarray}
\mu_p&=&\prod_{\alpha=0}^p\left(K\int
\prod_{n=1}^\infty dx^\alpha_n dx^\alpha_{-n}
\prod_{n=1}^\infty d\psi^\alpha_n d\psi^\alpha_{-n}
\,e^{-\half\sum_{n=1}^\infty x_{-n}^\alpha x_{n}^\alpha
-\half\sum_{n=1}^\infty\psi_{-n}^\alpha\psi_{n}^\alpha}
\right)\\
&=&\frac{1}{(2\pi)^{p+1}}.
\end{eqnarray}

\section{A-roof genus in CS-term}
\label{AppAroof}
Here we summarize the computation
to obtain the A-roof genus in the CS-term
explained in section \ref{index}.
We will show that if we include the scalar fields
and perform the path integral in the $u\ra\infty$ limit,
we will recover the CS-term of the form (\ref{indexg}),
where $R$ is replaced by the curvature two form with
respect to the induced metric.

When we include the scalar field, the boundary action
(\ref{curvedLag}) will become
\begin{eqnarray}
S_b&=& S_g+S_A+S_\phi\\
S_g&=&\int d\sigma\left(
\frac{1}{4u^2}\,
g_{\alpha\beta}(x)
\left(\dot x^\alpha \dot x^\beta+
\psi^\alpha\nabla_\sigma\psi^\beta
\right)\right),
\label{Sg}\\
S_A&=&\int d\sigma
\left(A_\alpha(x)\dot x^\alpha
-\half F_{\alpha\beta}(x)\psi^\alpha\psi^\beta
\right),\\
S_\phi&=&\int d\sigma\left(
iP_\alpha x^\alpha+i\Pi_\alpha \psi^\alpha
+iP_i\phi^i(x)
+i\Pi_i\del_\alpha\phi^i(x) \psi^\alpha\right).
\end{eqnarray}

We split the fields $x$ and $\psi$ into
zero modes $x_0$, $\psi_0$ and  non-zero modes $\delta x$, $\delta\psi$
as
\begin{eqnarray}
x=x_0+\delta x,~~~\psi=\psi_0+\delta \psi,
\end{eqnarray}
and consider the path integral with respect to
the non-zero modes.
Note that the $S_\phi$ will act on the boundary state as
\begin{eqnarray}
&&e^{-S_\phi}\ket{x^\mu=0}\ket{\psi^\mu=0}_{\rm RR}\\
&=&\ket{x^\alpha,x^i=\phi^i(x)}
\ket{\psi^\alpha,\psi^i=\del_\alpha\phi^i(x)\psi^\alpha}_{\rm RR}
\label{bch}\\
&=&
e^{
\sum_{m=1}^\infty
\left\{-\frac{1}{\alpha'}
\left(\ol g_{\alpha\beta}(x_0)
+\frac{1}{m}\ol R_{\alpha\beta}
(x_0)\right)x_{-m}^\alpha x_{m}^\beta
-\frac{1}{\alpha'}\ol g_{\alpha\beta}(x_0)
\psi_{-m}^\alpha\psi_{m}^\beta
+\cdots
\right\}}
\ket{x_0}\ket{\psi_0}_{\rm RR},
\label{gR}
\end{eqnarray}
where $\ol g_{\alpha\beta}(x_0)=\delta_{\alpha\beta}
+\del_\alpha\phi^i(x_0)\del_\beta\phi^i(x_0)$ is the induced metric
and $\ol R_{\alpha\beta}(x_0)=\half
\ol R_{\alpha\beta\gamma\delta}(x_0)\psi_0^\gamma\psi_0^\delta$
is the curvature two form defined by the induced metric.
Here we adopt the Riemann normal coordinates at $x_0$.
(\ref{gR}) is obtained by inserting the expansions like
\begin{eqnarray}
\phi^i(x)=\phi^i(x_0)+
\del_\alpha\phi^i(x_0)\delta x^\alpha+
\half\del_\alpha\del_\beta\phi^i(x_0)\delta x^\alpha\delta x^\beta+
\cdots
\end{eqnarray}
into (\ref{ketx}) and (\ref{ketpsi}).
In (\ref{gR}), we recovered the $\alpha'$ dependence
treating $\ol g_{\alpha\beta}$ as a dimensionless field.

Let us now consider the CS-term
 defined in (\ref{CS}) and
we would like to perform the path integral with respect to
the non-zero modes. We evaluate the Gaussian integral
coming from (\ref{gR}) and consider other terms
as perturbation. The contribution from
$S_g$ in (\ref{Sg}) can be dropped by taking the $u\ra\infty$ limit.
Then,
the Gaussian integral will become exact in the $\alpha'\ra 0$ limit.
As a result, we obtain the CS-term
\begin{eqnarray}
S_{CS}^{D(-1)}&=& C_0\,\mu_p
\int dx_0^\alpha d\psi_0^\alpha\,\Tr
e^{\pi F_{\alpha\beta}\psi_0^\alpha\psi_0^\beta}
\prod_{m=1}^\infty\frac{1}{\det\left(1-\frac{1}{m}\ol R\right)}\\
&=& C_0\int\Tr e^{F/2\pi}\wedge \wh A(\ol R)\label{CSAroof}
\end{eqnarray}
where $\ol R=(\ol R^\alpha_{~\beta})$.
To obtain the A-roof genus in (\ref{CSAroof}),
we have used the formula
\begin{eqnarray}
\prod_a\prod_{m=1}^\infty
\frac{1}{\left(1+\frac{(x_a/2\pi)^2}{m^2}\right)}
=\prod_a \frac{x_a/2}{\sinh \left(x_a/2 \right)}
=\wh A(\ol R),
\end{eqnarray}
where $x_a$'s are the skew eigen values of the curvature two-form
$\ol R/2\pi$.

Note also that if we consider $u\ra 0$ limit
with fixed $\alpha'$, we will again recover (\ref{indexg})
which is written in terms of the internal metric $g_{\alpha\beta}$.

Finally, we comment about the topological invariance
of the CS-term coupled to $C_0$.
Since the massless RR-state $\ket{C}$ does not contain
higher level oscillators, the oscillator dependent part
in (\ref{bch}) will drop in the computation of the CS-term.
Then, the relevant part of (\ref{gR}) can be rewritten using
superfields as
\beq
\exp\left\{-\frac{1}{4}
\int d\wh\sigma \left(
D \phi^i(\bx) \frac{1}{\sqrt{D^4}} D^2 \phi^i(\bx)
\right) \right\}
\ket{x_0}
\ket{\psi_0}_{\rm RR},
\eeq
where $D^2=\del_\tau$ and $\frac{1}{\sqrt{D^4}} D^2$
is well-defined non-local operator although
$\frac{1}{\sqrt{D^4}}$ is not.
Therefore, the coupling to the constant $C_0$ in the CS-term
becomes the partition function of
a supersymmetric theory, which can be rewritten as an index
of the supercharge.
This implies the topological invariance of the
D-instanton charge.
Note that the charge density is
not topological invariant as we have seen above.


\begin{thebibliography}{999}

\bibitem{Pol et al}
J.~Dai, R.~G.~Leigh and J.~Polchinski,
Mod.\ Phys.\ Lett.\ A {\bf 4} (1989) 2073.

\bibitem{CLNY}
C.~G.~Callan, C.~Lovelace, C.~R.~Nappi and S.~A.~Yost,
``Adding Holes And Crosscaps To The Superstring,''
Nucl.\ Phys.\ B {\bf 293} (1987) 83;
``Loop Corrections to Superstring Equations of Motion,''
Nucl.\ Phys.\ B {\bf 308} (1988) 221.

\bibitem{bdry}
For a review,
P. Di Vecchia and A. Liccardo,
``D-branes in string theory, I,''
hep-th/9912161.

\bibitem{BFSS}
T. Banks, W. Fischler, S.H. Shenker and L. Susskind,
``M Theory as a Matrix Model: A Conjecture,''
Phys. Rev. {\bf 55} (1997) 5112,
hep-th/9610043.

\bibitem{Sen}
A.~Sen,
``$SO(32)$ Spinors of Type I and Other Solitons
on Brane-Antibrane Pair,''
JHEP {\bf 9809} (1998) 023,
hep-th/9808141;\\
A.~Sen,
``Non-BPS states and branes in string theory,''
hep-th/9904207, and references therein.

\bibitem{MiMo}
R. Minasian and G. Moore,
``K-theory and Ramond-Ramond charge,''
JHEP {\bf 9711} (1997) 002,
hep-th/9710230.

\bibitem{WittenK}
E. Witten,
``D-branes and K-theory,'' JHEP {\bf 9812} (1998)
 019, hep-th/9810188.

\bibitem{Ho}
P. Horava,
``Type IIA D-Branes, K-Theory, and Matrix Theory,''
Adv. Theor. Math. Phys. {\bf 2} (1999) 1373, hep-th/9812135.

\bibitem{Te}
S. Terashima, ``A Construction of Commutative D-branes from Lower Dimensional
Non-BPS D-branes,''
JHEP {\bf 0105} (2001) 059,
hep-th/0101087.

\bibitem{Kl}
J.~Kluson,
``D-branes from N non-BPS D0-branes,''
JHEP {\bf 0011} (2000) 016,
hep-th/0009189.

\bibitem{AsSuTe}
T.~Asakawa, S.~Sugimoto and S.~Terashima,
``D-branes, matrix theory and K-homology,''
JHEP {\bf 0203} (2002) 034,
hep-th/0108085.

\bibitem{Halpern}
K. Bardakci, ``Dual Models and Spontaneous Symmetry Breaking,''
Nucl. Phys. {\bf B68} (1974) 331;
K. Bardakci and M. B. Halpern, ``Explicit Spontaneous Breakdown
in a Dual Model,''
Phys. Rev. {\bf D10} (1974) 4230;
K. Bardakci and M. B. Halpern, ``Explicit Spontaneous Breakdown
in a Dual Model II: N Point Functions,''
Nucl. Phys. {\bf B96} (1975) 285;
K. Bardakci, ``Spontaneous Symmetry Breakdown in the Standard Dual
String Model,''
Nucl. Phys. {\bf B133} (1978) 297;


\bibitem{Senbdry}
A.~Sen,
``BPS D-branes on non-supersymmetric cycles,''
JHEP {\bf 9812} (1998) 021,
hep-th/9812031. \\
A.~Sen,
``Descent relations among bosonic D-branes,''
Int.\ J.\ Mod.\ Phys.\ A {\bf 14} (1999) 4061,
hep-th/9902105.\\
J.~Majumder and A.~Sen,
``Blowing up D-branes on Non-supersymmetric Cycles,''
JHEP {\bf 9909} (1999) 004,
hep-th/9906109.\\
J.~Majumder and A.~Sen,
``Vortex pair creation on brane-antibrane pair via marginal deformation,''
JHEP {\bf 0006} (2000) 010,
hep-th/0003124.\\
M.~Naka, T.~Takayanagi and T.~Uesugi,
``Boundary state description of tachyon condensation,''
JHEP {\bf 0006} (2000) 007,
hep-th/0005114.

\bibitem{Is}
N. Ishibashi,
``$p$-branes from $(p-2)$-branes
in the Bosonic String Theory,''
Nucl. Phys. {\bf B539} (1999) 107,
hep-th/9804163.

\bibitem{WiBSFT}
E.~Witten,
``On background independent open string field theory,''
Phys.\ Rev.\ D {\bf 46} (1992) 5467,
hep-th/9208027.

\bibitem{GeSh}
A.~A.~Gerasimov and S.~L.~Shatashvili,
``On exact tachyon potential in open string field theory,''
JHEP {\bf 0010} (2000) 034,
hep-th/0009103.

\bibitem{KuMaMo}
D.~Kutasov, M.~Marino and G.~W.~Moore,
``Some exact results on tachyon condensation in string field theory,''
JHEP {\bf 0010} (2000) 045,
hep-th/0009148.

\bibitem{KuMaMo2}
D. Kutasov, M. Marino and G. Moore, ``Remarks on tachyon
condensation in superstring field theory,'' hep-th/0010108.

\bibitem{Ts2}
A.A. Tseytlin, ``Sigma model approach to string theory effective actions
with tachyons,''
J. Math. Phys. {\bf 42} (2001) 2854, hep-th/0011033.

\bibitem{KrLa}
P. Kraus and F. Larsen, ``Boundary String Field Theory of the DDbar
 System,'' Phys. Rev. {\bf D63} (2001) 106004, hep-th/0012198.

\bibitem{TaTeUe}
T. Takayanagi, S. Terashima and T. Uesugi, ``Brane-Antibrane Action from
Boundary String Field Theory,''
JHEP  {\bf 0003} (2001) 019, hep-th/0012210.

\bibitem{deAl}
S.P. de Alwis,
``Boundary String Field Theory, the Boundary State Formalism and
D-Brane Tension,''
Phys. Lett. {\bf B505} (2001) 215,
hep-th/0101200.


\bibitem{AnTs}
O.~D.~Andreev and A.~A.~Tseytlin,
``Partition Function Representation For The Open Superstring Effective Action:
Cancellation Of Mobius Infinities And Derivative Corrections To
Born-Infeld Lagrangian,''
Nucl.\ Phys.\ B {\bf 311} (1988) 205.

\bibitem{Connes}
A. Connes,
``Noncommutative geometry,'' Academic Press, 1994.\\
See also,\\
A. Connes, ``A Short Survey of Noncommutative Geometry,''
hep-th/0003006.\\
A. Connes, ``Noncommutative Geometry Year 2000,''
math.QA/0011193.

\bibitem{AsSuTe2}
T.~Asakawa, S.~Sugimoto and S.~Terashima,
``D-branes and KK-theory in type I string theory,''
JHEP {\bf 0205} (2002) 007,
hep-th/0202165.

\bibitem{HaKuMa}
J.~A.~Harvey, D.~Kutasov and E.~J.~Martinec,
``On the relevance of tachyons,''
hep-th/0003101.

\bibitem{Wy}
N. Wyllard,
``Derivative corrections to D-brane actions
with constant background fields,''
Nucl. Phys. {\bf B598} (2001) 247,
hep-th/0008125.

\bibitem{DiVeFrLeLi}
P.~Di Vecchia, M.~Frau, A.~Lerda, A. Liccardo,
``(F,D$p$) bound states from the boundary state.''
Nucl.\ Phys.\ B {\bf 565} (2000) 397,
hep-th/9906214.

\bibitem{TeUe}
S.~Terashima and T.~Uesugi, ``On the Supersymmetry of
Non-BPS D-brane,''
JHEP {\bf 0105} (2001) 054, hep-th/0104176.

\bibitem{OkOo}
Y. Okawa and H. Ooguri,
``An Exact Solution to Seiberg-Witten Equation of Noncommutative
Gauge Theory,''
Phys. Rev. {\bf D64} (2001) 046009,
hep-th/0104036.

\bibitem{MuSu}
S.~Mukhi and N.~V.~Suryanarayana,
``Gauge-invariant couplings of noncommutative branes to Ramond-Ramond
 backgrounds,''
JHEP {\bf 0105} (2001) 023,
hep-th/0104045.

\bibitem{LiMi}
H.~Liu and J.~Michelson,
``Ramond-Ramond couplings of noncommutative D-branes,''
Phys.\ Lett.\ B {\bf 518} (2001) 143,
hep-th/0104139.


\bibitem{DiVecchia et al}
M.~Billo, P.~Di Vecchia, M.~Frau, A.~Lerda, I.~Pesando, R.~Russo and S.~Sciuto,
``Microscopic string analysis of the D0-D8 brane system and dual R-R  states,''
Nucl.\ Phys.\ B {\bf 526} (1998) 199,
hep-th/9802088.

\bibitem{KeWi}
C.~Kennedy and A.~Wilkins,
Phys.\ Lett.\ B {\bf 464} (1999) 206,
hep-th/9905195.

\bibitem{Qu}D. Quillen, ``Superconnection and the Chern character,"
Topology {\bf 24} (1985) 89.

\bibitem{My}
R.C. Myers,
``Dielectric-Branes,''
JHEP {\bf 9912} (1999) 022,
hep-th/9910053.

\bibitem{Atiyah-Singer}
M. F. Atiyah and I. M. Singer,
``The index of elliptic operators I, III,''
Ann. of Math. 87 (1968), 484-530, 546-604.

\bibitem{Al}
L.~Alvarez-Gaume,
``Supersymmetry And The Atiyah-Singer Index Theorem,''
Commun.\ Math.\ Phys.\  {\bf 90} (1983) 161.

\bibitem{FrWi}
D.~Friedan and P.~Windey,
``Supersymmetric Derivation Of The Atiyah-Singer Index
And The Chiral Anomaly,''
Nucl.\ Phys.\ B {\bf 235} (1984) 395.

\bibitem{CS-term}
M.~B.~Green, J.~A.~Harvey and G.~W.~Moore,
``I-brane inflow and anomalous couplings on D-branes,''
Class.\ Quant.\ Grav.\  {\bf 14} (1997) 47,
hep-th/9605033.\\
Y.~K.~Cheung and Z.~Yin,
``Anomalies, branes, and currents,''
Nucl.\ Phys.\ B {\bf 517} (1998) 69,
hep-th/9710206.

\bibitem{AlSc}
A.~Alekseev and V.~Schomerus,
``RR charges of D2-branes in the WZW model,''
hep-th/0007096.

\bibitem{Is2}
N.~Ishibashi,
``A relation between commutative and noncommutative
descriptions of  D-branes,''
hep-th/9909176.

\bibitem{Co}
L.~Cornalba,
``D-brane physics and noncommutative Yang-Mills theory,''
Adv.\ Theor.\ Math.\ Phys.\  {\bf 4} (2000) 271,
hep-th/9909081.

\bibitem{Ok}
K. Okuyama,
``Boundary States in B-Field Background,''
Phys. Lett. {\bf B499} (2001) 305,
hep-th/0009215.

\bibitem{Ok2}
K. Okuyama,
``A Path Integral Representation of the Map between
Commutative and Noncommutative Gauge Fields,''
JHEP {\bf 0003} (2000) 016,
hep-th/9910138.

\bibitem{SW}
N.~Seiberg and E.~Witten,
JHEP {\bf 9909} (1999) 032,
hep-th/9908142.

\bibitem{AkGeSh}
E.~T.~Akhmedov, A.~A.~Gerasimov and S.~L.~Shatashvili,
``On unification of RR couplings,''
JHEP {\bf 0107} (2001) 040,
hep-th/0105228.

\bibitem{FGLS}
M. Frau, L. Gallot, A. Lerda and P. Strigazzi,
''Stable non-BPS D-Branes in Type I String Theory,''
Nucl. Phys. {\bf B564} (2000) 60,
hep-th/9903123;\\
M. Frau, L. Gallot, A. Lerda and P. Strigazzi,
''Stable non-BPS D-Branes of Type I,''
hep-th/0003022.

\bibitem{bergman}
O.~Bergman,
``Tachyon condensation in unstable type I D-brane systems,''
JHEP {\bf 0011} (2000) 015,
hep-th/0009252.

\bibitem{SuTe}
S.~Sugimoto and S.~Terashima,
``Tachyon matter in boundary string field theory,''
JHEP {\bf 0207} (2002) 025,
hep-th/0205085.

\bibitem{Mi}
J.~A.~Minahan,
``Rolling the tachyon in super BSFT,''
JHEP {\bf 0207} (2002) 030,
hep-th/0205098.

\bibitem{Yi}
P.~Yi,
``Membranes from five-branes and fundamental strings from Dp branes,''
Nucl.\ Phys.\ B {\bf 550} (1999) 214,
hep-th/9901159.

\bibitem{BeHoYi}
O.~Bergman, K.~Hori and P.~Yi,
``Confinement on the brane,''
Nucl.\ Phys.\ B {\bf 580} (2000) 289,
hep-th/0002223.

\bibitem{HaKrLaMa}
J.~A.~Harvey, P.~Kraus, F.~Larsen and E.~J.~Martinec,
``D-branes and strings as non-commutative solitons,''
JHEP {\bf 0007} (2000) 042,
hep-th/0005031.

\bibitem{SenN}A. Sen, ``Supersymmetric World-volume Action for Non-BPS
 D-branes,'' JHEP {\bf 9910} (1999) 008, hep-th/9909062.

\bibitem{Yo}
T. Yoneya, ``Spontaneously Broken Space-Time Supersymmetry in Open String
Theory without GSO Projection,''
Nucl.\ Phys.\ {\bf B576} (2000) 219, hep-th/9912255.

\bibitem{LiWi}
K.~Li and E.~Witten,
``Role of short distance behavior in off-shell open string field theory,''
Phys.\ Rev.\ D {\bf 48} (1993) 853,
hep-th/9303067.

\bibitem{AIIKKT}
H. Aoki, N. Ishibashi, S. Iso, H. Kawai, Y. Kitazawa and T. Tada,
``Noncommutative Yang-Mills in IIB Matrix Model,''
Nucl. Phys. {\bf B 565} (2000) 176,
hep-th/9908141.

\bibitem{Marcus}
N.~Marcus,
``Open String And Superstring Sigma Models With Boundary Fermions,''
DOE-ER-40423-09-P8.


\end{thebibliography}
\end{document}